\documentclass[sigconf, nonacm]{acmart}

\usepackage{hyperref}
\usepackage{url}
\usepackage[utf8]{inputenc}
\usepackage[T1]{fontenc}
\usepackage{amsmath}
\usepackage{amsthm}
\usepackage{amsfonts}
\usepackage{booktabs}
\usepackage{multirow}
\usepackage{adjustbox}
\usepackage{rotating}
\usepackage{thm-restate}
\usepackage{enumitem}
\usepackage{subcaption}

\usepackage[linesnumbered, ruled, vlined]{algorithm2e}
\DontPrintSemicolon
\SetKwInOut{Input}{Input}
\SetKwInOut{Output}{Output}
\SetKwProg{function}{Function}{:}{}
\SetKwProg{parfor}{parallel for}{\:do}{}

\SetVlineSkip{0.2pt}        
\SetInd{0.7em}{0.7em}     

\usepackage{graphicx}
\usepackage{enumitem}
\usepackage{textgreek}
\usepackage{tabularx}
\usepackage{xspace}
\PassOptionsToPackage{dvipsnames,svgnames}{xcolor}
\usepackage{tcolorbox}

\newtheorem{theorem}{Theorem}
\newtheorem{proposition}{Proposition}

\newtheorem{claim}[theorem]{Claim}
\newtheorem{definition}{Definition}

\makeatletter
\newcommand{\multiline}[1]{%
  \begin{tabularx}{\dimexpr\linewidth-\ALG@thistlm}[t]{@{}X@{}}
    #1
  \end{tabularx}
}
\makeatother



\def \cC {\mathcal{C}\xspace}

\def \cL {\mathcal{L}\xspace}

\def \cP {\mathcal{P}\xspace}

\DeclareMathOperator*{\argmin}{arg\,min}

\newcommand{\R}{\ensuremath{\mathbb{R}}}

\newcommand{\myparagraph}[1]{\vspace{.05in}\noindent {\bf #1.}}

\newcommand{\flatten}{\textsf{Flatten}\xspace}
\newcommand{\mvlloyd}{\textrm{MV-Lloyd}\xspace}
\newcommand{\mvivf}{\textrm{MV-IVF}\xspace}
\newcommand{\mvvamana}{\textrm{MV-Vamana}\xspace}
\newcommand{\mvsic}{\textrm{MVSIC}\xspace}
\newcommand{\Ch}{\ensuremath{\mathtt{Ch}}}
\newcommand{\Chn}{\ensuremath{\mathtt{Ch}}}

\newcommand{\cost}{\ensuremath{\mathtt{cost}}}
\newcommand{\maxsim}{\mathrm{MaxSim}\xspace}

\newcommand{\leafsize}{\ell_{\mathrm{\max}}\xspace}
\newcommand{\branch}{b\xspace}
\newcommand{\centersize}{s\xspace}
\newcommand{\nprobe}{\texttt{nprobe}\xspace}
\newcommand{\beamsize}{\texttt{beam\_size}\xspace}

\newcounter{mnote}[section]
\newcommand{\thenote}{\thesection.\arabic{mnote}}

\newcommand{\kishen}[1]{\refstepcounter{mnote}$\ll${\bf Kishen~\thenote:}{\sf \color{violet} #1}$\gg$\marginpar{\tiny\bf KNG~\thenote}}

\newcommand{\evalfigure}[5]{%
    \IfFileExists{plots/#1}{%
        \begin{figure}[!htbp]
            \centering
            \includegraphics[width=\linewidth]{plots/#1}
            \caption{\small \textbf{#2 on \texttt{#3}.} #4}
            \label{#5}
        \end{figure}
    }{}%
}

\newcommand{\supponly}[1]{%
  #1%
}

\begin{document}
\title{MV-IVF: Multi-Vector Retrieval via Multi-Vector Clustering}

\author{Kishen N Gowda}
\affiliation{%
  \institution{University of Maryland}
  \city{College Park}
  \state{MD}
  \country{USA}
}

\author{Laxman Dhulipala}
\affiliation{%
  \institution{University of Maryland}
  \city{College Park}
  \state{MD}
  \country{USA}
}

\author{Lars Gottesb{\"u}ren}
\affiliation{%
  \institution{Google Research}
  \city{Z{\"u}rich}
  \country{Switzerland}
}

\author{Rajesh Jayaram}
\affiliation{%
  \institution{Google Research}
  \city{New York City}
  \state{NY}
  \country{USA}
}

\begin{abstract}
Multi-vector retrieval represents each query and document as a set of embeddings and scores relevance via fine-grained set-to-set interactions such as \emph{MaxSim} (Chamfer similarity). This expressiveness improves retrieval quality over single-vector representations, but makes search substantially more expensive. Existing systems address this challenge primarily through proxy representations or filter-and-rerank pipelines that reduce candidate generation to single-vector search. In this work, we study the complementary direction of building a native IVF-style index directly over multi-vector data.

We introduce \emph{MV-IVF}, an inverted-file index for multi-vector retrieval built on \emph{MV-Lloyd}, a new Lloyd-style algorithm for clustering point clouds under a Chamfer objective. Our key observation is that once point clouds are assigned to clusters, the center-update step reduces to a standard single-vector clustering problem over the flattened embeddings in each cluster. This enables efficient construction of a hierarchical IVF tree whose internal nodes are represented by center point clouds.

At query time, MV-IVF performs best-first beam search over the tree to identify promising leaf clusters, then scores the documents in those leaves. We further accelerate search using low-bit quantization, query compression, and cache-efficient batched search. We implement MV-IVF as part of \emph{MVSIC}, an optimized C++ library and unified evaluation framework for multi-vector retrieval that includes several major multi-vector retrieval approaches. Across BEIR and LoTTE text retrieval benchmarks, MV-IVF achieves strong latency--recall tradeoffs, the best batched throughput frontiers, and favorable build-time and index-size profiles. At $90\%$ Recall-$10$@$10$, MV-IVF achieves up to $8.03\times$ lower single-query latency and up to $173.5\times$ higher QPS than the baselines, demonstrating that native IVF-style indexing is a practical and effective approach for scalable multi-vector retrieval.
\end{abstract}

\maketitle


\section{Introduction}\label{sec:intro}

Multi-vector retrieval has emerged as a powerful paradigm for modern retrieval systems, in which each query and document is represented not by a single embedding, but by a set of embeddings (i.e., a point cloud). 
Such representations arise naturally in late-interaction models for text retrieval (e.g., ColBERT~\cite{khattab2020colbert,santhanam2022colbertv2}), where relevance is computed via token-level interactions using operators such as $\maxsim$, as well as in region-based representations for images and other modalities (e.g., ColPali~\cite{faysse2025colpali}). This added expressiveness yields significantly better retrieval quality than single-vector encoders (e.g., on standard benchmarks such as BEIR~\cite{thakur2021beir}, LoTTE~\cite{santhanam2022colbertv2}, and ViDoRe~\cite{faysse2025colpali}) but the increased complexity of the representation makes search substantially more expensive.

The gold-standard multi-vector similarity, introduced by ColBERT~\cite{khattab2020colbert} and used by most subsequent late-interaction models, is $\maxsim$. Given a query point cloud $A \subseteq \R^d$ and a document point cloud $B \subseteq \R^d$,
$$\maxsim(A,B) := \sum_{a \in A} \max_{b \in B} \langle a, b \rangle,$$ 
i.e., each query embedding is matched to its most similar document embedding and the resulting scores are summed. Thus, unlike single-vector retrieval, even a single query--document $\maxsim$ score can require many inner product computations. At scale, the central challenge is therefore to recover the top-$k$ documents under $\maxsim$ without exhaustively scoring the entire collection.

A large body of recent work has focused on making multi-vector retrieval under $\maxsim$ efficient. Most practical approaches follow a common philosophy: rather than indexing point clouds directly, they reduce candidate generation to a simpler proxy or filtering problem built using single-vector retrieval techniques, and reserve exact multi-vector scoring for a smaller set of candidates. For example, PLAID~\cite{santhanam2022plaid} and its follow-up, WARP~\cite{scheerer2025warp}, use single-vector search over clustered token centroids to generate candidates, which are then scored using quantized and exact $\maxsim$; \citet{dhulipala2024muvera} formalize this style of pipeline as the \emph{single-vector heuristic} (SVH). MUVERA~\cite{dhulipala2024muvera} instead constructs fixed-dimensional encodings (FDEs), large single-vector representations whose inner products provably approximate multi-vector similarity, thereby reducing the problem to standard single-vector MIPS. These approaches are highly effective, but the quality of the final top-$k$ results depends on how well the surrogate representation or candidate-generation stage preserves documents that rank highly under $\maxsim$.

This motivates a complementary direction: \emph{can we build scalable indexes that operate directly over multi-vector representations while accurately preserving the top-$k$ results under $\maxsim$?} A smaller body of work has explored such native multi-vector indexing through hashing- and graph-based approaches, including DESSERT~\cite{engels2023dessert} and IGP~\cite{bian2025igp}. In this work, we investigate a different direction: extending clustering-based inverted-file (IVF) indexing, one of the most widely used paradigms for large-scale single-vector retrieval~\cite{jegou2011pq,douze2024faiss}, to multi-vector data.

IVF is an appealing candidate for this setting. In the single-vector case, an IVF index partitions the dataset into clusters and probes only those whose representatives are most relevant to the query. This organization is simple and scalable to build, has low indexing overhead, exposes substantial parallelism, and is naturally suited to high-throughput batched search. These properties are particularly attractive for multi-vector retrieval, where scoring each candidate is substantially more expensive. However, extending IVF to multi-vector data is non-trivial: \emph{how should we cluster point clouds, and what representatives should we use to route queries to the resulting clusters?}

We address these challenges through a set of contributions centered around \mvivf, our native IVF-style index for multi-vector retrieval.

\subsection{Our Contributions}
Our main contribution is \textbf{\mvivf}, a hierarchical IVF-style index for multi-vector retrieval. Our approach represents each cluster by a center that is itself a point cloud containing at most $s$ vectors, and uses $\maxsim$ scores between the query and these center point clouds for routing. In this sense, \mvivf is a \emph{native} multi-vector index: both the indexed objects and the routing representatives remain point clouds, rather than being reduced to single-vector surrogates.

The key algorithmic ingredient is \textbf{\mvlloyd}. We first formulate a natural multi-vector analogue of $k$-median, in which each cluster center is a point cloud of size at most $s$, and assignment costs are measured using the analogous Chamfer distance:
$$
  \Ch(A,B) := \frac{1}{|A|} \sum_{a \in A} \min_{b\in B} \|a-b\|_2.
$$
We then develop \mvlloyd, a practical Lloyd-style algorithm for this problem. Our key observation is that, once the point-cloud assignments to clusters are fixed, finding the optimal center point cloud for a cluster reduces exactly to a weighted $s$-median problem over the flattened embeddings of the point clouds assigned to that cluster. This allows the center-update step to leverage standard, highly optimized single-vector clustering machinery.

\mvivf recursively applies \mvlloyd to construct a tree-structured partition of the document point clouds and stores the resulting center point clouds at internal nodes for routing. At query time, \mvivf performs a best-first beam search over the tree to identify a small number of promising leaf clusters and scores only the documents contained in those leaves. We further accelerate search using low-bit quantization, query compression, and early termination. 

For high-throughput workloads, many queries probe overlapping sets of leaves, creating an opportunity to reuse leaf data across queries. Rather than processing each query independently, we group queries that probe the same leaf and score them together, allowing the leaf data to be reused while cache-resident. This is particularly valuable in the multi-vector setting, where each leaf contains many document point clouds, each comprising many embeddings, and repeatedly loading the same leaf can incur substantial memory traffic.

To further explore the design space of native multi-vector indexes, we also develop \textbf{\mvvamana}, a graph-based index extending the Vamana algorithm~\cite{subramanya2019diskann} to operate directly over point clouds. \mvvamana builds a proximity graph over document point clouds and performs beam search using multi-vector scores directly. We find that it is highly competitive in the single-query regime, where graph-based search is effective at minimizing the number of expensive point-cloud comparisons.

Finally, we develop \textbf{M}ulti-\textbf{V}ector \textbf{S}earch, \textbf{I}ndexing and \textbf{C}lustering (\textbf{MVSIC}, pronounced ``music''), an optimized C++ library and unified evaluation framework for multi-vector retrieval. MVSIC includes implementations of SVH, MUVERA, \mvvamana, and \mvivf using a common framework, optimized scoring kernels, and evaluation pipeline, and also supports direct comparison with PLAID and IGP.

We evaluate MVSIC on eight text retrieval datasets from BEIR and the pooled LoTTE benchmark, using ColBERTv2 multi-vector embeddings. We measure retrieval quality against the exact $\maxsim$ ranking, with particular emphasis on Recall-$k$@$k$, which measures how accurately the full exact top-$k$ neighborhood is recovered.

Across these datasets, \mvivf achieves strong latency--recall tradeoffs and the best batched throughput frontiers across recall metrics. At $90\%$ Recall-$10$@$10$, \mvivf achieves up to {\boldmath$8.03\times$} lower single-query latency and up to {\boldmath$173.5\times$} higher QPS than MUVERA, SVH, and \mvvamana; the corresponding geometric-mean improvements are {\boldmath$1.22$--$1.55\times$} in latency and {\boldmath$2.07$--$4.01\times$} in throughput. \mvivf also has a favorable build-time and index-size profile, with practical construction times and moderate index size. \mvvamana is highly competitive in the single-query regime, although its construction is substantially more expensive. We additionally compare against FastPLAID and IGP on \texttt{arguana} and \texttt{fiqa}, where \mvivf achieves substantially higher batched throughput, including over one and two orders of magnitude over IGP and FastPLAID, respectively.

Taken together, our results show that native IVF-style indexing provides a practical route to scalable, high-quality multi-vector retrieval: \mvivf accurately recovers the top-$k$ neighborhood under $\maxsim$, supports both low-latency and high-throughput search, and retains the practical build-time and space properties that make IVF attractive in the single-vector setting. To the best of our knowledge, MVSIC provides the most comprehensive evaluation of multi-vector retrieval methods across a wide range of datasets to date. 


\section{Background}\label{sec:background}
\myparagraph{Multi-vector retrieval}
In multi-vector retrieval, each query and each document is represented as a \emph{set} of embeddings, or equivalently, a point cloud.
Formally, let $\cP = \{P_1,\ldots,P_n\}$ denote a collection of document point clouds, where each $P_i \subseteq \R^d$ is a finite set of embeddings, and let $Q \subseteq \R^d$ denote a query point cloud.
The retrieval task is to return the top-$k$ document point clouds most relevant to $Q$ under a suitable set-to-set scoring rule.

\begin{definition}[Chamfer distance and MaxSim]
Let $A,B \subseteq \R^d$ be finite sets of embeddings. The (asymmetric) Chamfer distance from $A$ to $B$ is 
$$\Ch(A,B) := \frac{1}{|A|}\sum_{a \in A} \min_{b \in B} \|a-b\|_2.$$
The $\maxsim$ similarity is
$$\maxsim(A,B) := \sum_{a \in A} \max_{b \in B}  \langle a,b\rangle.$$
\end{definition}

Multi-vector retrieval is typically phrased in terms of $\maxsim$: throughout the paper, we use $\maxsim$ to rank documents and to route queries during search.
For the clustering primitive underlying our IVF construction (defined and studied in Section~\ref{sec:mvclustering}), it is more natural to work with the Chamfer distance. 

\myparagraph{Inverted-file (IVF) indexes}
In the single-vector setting, an inverted-file (IVF) index partitions the documents into clusters and stores the points assigned to each cluster in an inverted list. Each cluster is represented by a centroid; at query time, the query is compared against these centroids and only a small number of promising lists are probed. Hierarchical variants recursively partition the data, yielding a tree whose internal nodes store routing representatives and whose leaves contain the database points. Sections~\ref{sec:mvclustering} and~\ref{sec:mvivf} develop the corresponding clustering and indexing primitives for multi-vector data.


\section{Multi-Vector Clustering}\label{sec:mvclustering}

In this section, we develop \mvlloyd, a practical Lloyd-style heuristic for clustering point clouds under a Chamfer-based objective. We now formally define the multi-vector clustering problem.

\subsection{The Clustering Objective}
Let $\cP = \{P_1,P_2,\ldots,P_n\}$ denote a collection of point clouds, where each $P_i \subseteq \R^d$ is a finite set of embeddings. 
For clustering, we seek $k$ representative center-sets, each of size at most $s$, that minimize the total assignment cost under Chamfer distance.

\begin{definition}[Chamfer $(k,s)$-median]
Given $\cP = \{P_1,\ldots,P_n\}$ and parameters $k,s \in \mathbb{N}$, the goal is to find a collection of center-sets $\cC = \{C_1,C_2,\ldots,C_k\}$, $C_j \subseteq \R^d$, $|C_j|\le s$, minimizing

$$\cost(\cC) := \sum_{i=1}^n \min_{j\in[k]} \Chn(P_i,C_j).$$
\end{definition}

This objective can be viewed as a natural multi-vector analogue of the classical $k$-median clustering problem. 
Each center is itself a point cloud, and the budget $s$ controls the size, and hence the expressiveness, of the representative stored for each cluster. Larger values of $s$ allow the center-set to capture more of the structure of its assigned point clouds; if $s$ is large enough, the center-set can contain the union of all embeddings in the cluster.

\subsection{\mvlloyd}

We now present \mvlloyd, a practical Lloyd-style heuristic for iteratively refining a clustering of point clouds under the Chamfer $(k,s)$-median objective. As in classical Lloyd's algorithm, \mvlloyd alternates between two steps:

\begin{enumerate}[topsep=0pt,itemsep=0pt,parsep=1pt,leftmargin=20pt]
    \item \textbf{Assignment step:} assign each point cloud to its nearest center-set under Chamfer distance.
    \item \textbf{Update step:} recompute each center-set from the point clouds assigned to it.
\end{enumerate}
More concretely, given current centers $\cC = \{C_1,\ldots,C_k\}$, the assignment step induces a partition
\begin{align*}
    \Pi_j := \{P_i \in \cP : j = \argmin_{\ell\in[k]} \Chn(P_i,C_\ell)\},\; j\in[k].
\end{align*}
The main challenge in the multi-vector setting lies in the update step. Unlike vanilla $k$-means, where the optimal update is given by the arithmetic mean of the assigned vectors, here each updated center must itself be a point cloud of size at most $s$. Given a cluster $\Pi_j$, the new center is found by solving
\begin{align*}
    C_j = \argmin_{C \subseteq \R^d,\ |C|\le s} \sum_{P\in \Pi_j} \Chn(P,C).
\end{align*}
This is precisely a Chamfer $(1,s)$-median problem on the point clouds assigned to cluster $j$.

\IncMargin{1.2em}
\begin{algorithm}[t]
    \caption{\mvlloyd}\label{alg:mvlloyd}
    \Input{Collection of point clouds $\cP$, $k, s > 0$}
    \Output{Collection of $k$ point clouds $\cC = \{C_1,C_2,\ldots,C_k\}$, where $|C_i|=s$}
    \tcp{Step 1: Initialization}
    Let $\cC \gets $ uniform random sample of $k$ point clouds from $\mathcal{P}$\;
    \tcp{Step 2: Lloyd's Iterations}
    \For{$t=1$ to $T$}{
        \tcp{Assign point clouds to nearest centers}
        \parfor{$i=1$ to $k$}{
            $\Pi(i) \gets \{P\in \cP: i=\argmin_{j\in [k]}\Chn(P,\cC(j))\}$\;
        }

        \tcp{Compute new centroid point clouds}
        \parfor{$i=1$ to $k$}{
            $\cC(i) \gets \textsc{kmeans}(\flatten(\Pi(i)),s)$\;
        }
    }
    \Return{$\cC$}
\end{algorithm}
\DecMargin{1.2em}

\myparagraph{Center Updates via $s$-Median on the Flattened Set}
The key observation here is that the Chamfer $(1,s)$-median problem decomposes into a standard clustering problem on the \emph{individual points} appearing in the assigned point clouds. Specifically, for a cluster $\Pi_j \subseteq \cP$, let
$\flatten(\Pi_j) := \biguplus_{P\in \Pi_j} P$
denote the multiset obtained by flattening all points in the point clouds assigned to $\Pi_j$. 

\begin{proposition}\label{prp:mvlloyd}
For any cluster $\Pi_j$ and any candidate center-set $C \subseteq \R^d$ with $|C|\le s$,
\begin{align*}
    \sum_{P\in \Pi_j} \Chn(P,C) = \sum_{x\in \flatten(\Pi_j)} w(x)\cdot\min_{c\in C} d(x,c),
\end{align*}
where $w(x) = 1/|P|$ for $x\in P$.
Computing the optimal Chamfer $(1,s)$-median center-set for $\Pi_j$ is equivalent to solving a (weighted) $s$-median problem on the multiset $\flatten(\Pi_j)$. 
\end{proposition}

Proposition~\ref{prp:mvlloyd} shows that the center-update step has a particularly simple structure: once the assignment of point clouds to clusters is fixed, updating each center-set reduces to an ordinary weighted single-vector clustering problem on the flattened embeddings in that cluster. Thus, the multi-vector nature of the problem appears only in the assignment step; the update step can reuse standard single-vector clustering routines.

In the formal objective above, the update corresponds to a weighted \textbf{$s$-median} problem. 
In our implementation, however, we use an \textbf{$s$-means} update with cosine similarity, matching the geometry of the normalized embeddings used by the retrieval model and allowing us to leverage highly optimized $k$-means implementations. 
This gives a practical approximation to the ideal Chamfer center update.
Algorithm~\ref{alg:mvlloyd} summarizes the overall procedure.


\section{Multi-Vector IVF Index (MV-IVF)}\label{sec:mvivf}
In this section, we describe \mvivf, a clustering-based retrieval method that generalizes the inverted-file (IVF) framework to multi-vector data. 
During indexing, the data point clouds are recursively partitioned using \mvlloyd, producing a tree-structured IVF index whose internal nodes are represented by centroid point clouds. 
During search, a query is routed to relevant leaf clusters, and the document point clouds contained in those leaves are scored to obtain the final candidates.

The key distinction from standard single-vector IVF is that both the indexed objects and the routing representatives are point clouds. 
Thus, routing decisions are made directly in the native multi-vector space, using $\maxsim$ between the query point cloud and the centroid point clouds stored in the tree.

\subsection{Index Construction}
Given the point clouds $\cP$ associated with a node $u$, the construction proceeds as follows. 
If $|\cP| \leq \leafsize$, where $\leafsize$ is the maximum leaf size, then $u$ becomes a leaf and stores $\cP$. 
Otherwise, we run \mvlloyd on $\cP$ with branching factor $\branch$ and centroid point cloud size $\centersize$, obtaining a partition $\Pi_1,\Pi_2,\ldots,\Pi_\branch$ together with centroid point clouds $C_1,C_2,\ldots,C_\branch$.
We create one child for each cluster $\Pi_i$, store the corresponding centroid point cloud $C_i$ at that child, and recurse on $\Pi_i$.

In this way, \mvivf builds a tree whose leaves contain document point clouds, while the internal node store centroid point clouds used for routing. 
The resulting structure is a direct multi-vector analogue of a standard $k$-means tree, except that clustering and routing are done over point clouds.

\IncMargin{1.2em}
\begin{algorithm}[t]
    \caption{\textsc{GreedySearch}}
    \label{alg:greedy-search}
    \Input{Query point cloud $Q$, MVIVF tree $T$, \nprobe, and \beamsize}
    \Output{Top \texttt{nprobe} leaf nodes to probe}

    Initialize a \emph{beam} with the root node of $T$\;
    Initialize an empty top-$k$ structure $\cL$ with $k = \nprobe$\;

    \While{the beam is non-empty}{
        Extract the best node $u$ from the beam\;

        \If{$|\cL|=\nprobe$ ~and~ $\maxsim(Q,u)\le \min_{v\in \cL} \maxsim(Q,v)$}{
            \textbf{break} \tcp*[r]{early termination}
        }

        Compute $\maxsim$ values from $Q$ to all children of $u$\;

        \ForEach{child $v$ of $u$}{
            \eIf{$v$ is a leaf}{
                insert $v$ into $\cL$ if it belongs to the current top \nprobe leaves\;
            }{
                insert $v$ into the beam if it belongs to the current top \beamsize internal nodes\;
            }
        }
    }
    \Return{$\cL$}
\end{algorithm}
\DecMargin{1.2em}

\subsection{Search}
To identify the top \nprobe leafs to probe for a query point cloud $Q$, we perform a greedy beam search over the tree, prioritizing nodes by their $\maxsim$ score
with $Q$. The search maintains a beam of active internal nodes, truncated to
size \beamsize, and a list $\cL$ of the best \nprobe leaves encountered so far.
At each step, it expands the highest-scoring internal node, scores its child
centroid point clouds, and inserts the most promising child nodes into the
appropriate collection: internal nodes into the beam, and leaves into $\cL$.

We also employ the following early termination strategy. Suppose the algorithm has already found \nprobe leaf nodes. If the best remaining internal node in the beam has $\maxsim$ value no larger than the worst leaf currently in $\cL$, then the search terminates. Intuitively, this indicates that the remaining frontier is unlikely to contain a better probe leaf, and further exploration is unlikely to improve the final probe set. We note that this rule is a heuristic: the centroid point cloud stored at an internal node is used as a routing representative, rather than a certified upper bound on the scores of all leaves in its subtree. Empirically, however, this early termination substantially reduces search time with little effect on retrieval quality.


Algorithm~\ref{alg:greedy-search} summarizes the greedy search procedure. Once the top \nprobe leaves have been identified, \mvivf scores the document point clouds contained in those leaves using $\maxsim$ and returns the top-$k$ results overall.

\subsection{Batch Search}
In high-throughput settings, queries either arrive in, or can be readily grouped into \emph{batches}. A naive approach for this batch setting is to simply run the single-query algorithm independently per query, but this misses significant optimization opportunities as many queries end up probing the same leaf clusters, and repeatedly loading the same leaf data, wasting memory bandwidth. This structure is particular to IVF-style indexes: unlike graph-based beam search, whose traversal is irregular and query-dependent, IVF exposes shared leaf-level work that can be batched and scored with good locality.

To exploit this structure, \mvivf uses a batched search procedure that preserves the logical behavior of the single-query algorithm, but reorganizes the work to improve locality and vectorization. The input is a batch of query point clouds $Q_1,Q_2,\ldots,Q_m$, where each query may have its own search parameters such as $k$ and \nprobe. For each query, the goal is still to identify a set of probe leaves, score the document point clouds in those leaves, and return the final results.
The difference is that \mvivf schedules the expensive scoring work across queries so that shared tree and leaf data can be reused while it is cache-resident.

\myparagraph{Batched probe generation}
At the root, every query must score the same set of child centroid point clouds.
We therefore compute these scores for the entire query batch together, turning root scoring into a many-to-many $\maxsim$ computation.
After the root, each query continues its own greedy search independently.
Although the same batching idea could be applied deeper in the tree, doing so requires additional synchronization and gives diminishing returns as queries quickly diverge to different subtrees.

\myparagraph{Leaf-grouped scoring}
After probe generation, each query has a set of leaves that it needs to scan. Rather than processing each query--leaf pair independently, \mvivf groups the work by leaf. For each leaf $u$, we collect all queries whose probe set contains $u$, load the document point clouds stored in $u$, and score them against all queries assigned to that leaf. This improves cache locality by reusing the same leaf data across multiple queries and avoids repeatedly streaming large point-cloud data from memory.

\myparagraph{Gathering results}
The leaf-grouped organization introduces a final gathering step: after each leaf has scored the queries assigned to it, the resulting candidates must be collected back into per-query result sets.
We implement this by having each leaf-scoring kernel produce local top candidates for its assigned queries, then scatter these candidates into per-query buffers. Each query then aggregates the candidates returned from all of its probed leaves, keeps the best $k$ candidates, and returns them as the final top-$k$ results. This scatter-and-gather step adds synchronization and memory traffic, but in our setting this overhead is outweighed by the improved locality from grouping leaf scans and reusing leaf data across multiple queries.

The common computational primitive in both batched probe generation and leaf-grouped scoring is a many-to-many $\maxsim$ problem: a group of query point clouds must be scored against a group of document point clouds stored in the same leaf. For exact scoring, we decompose this into a sequence of vectorized matrix multiplications and reduction operations, using optimized linear-algebra kernels where appropriate. We also use standard systems techniques such as cache-aware blocking and software prefetching to reduce memory stalls during leaf scans. Nevertheless, exact many-to-many $\maxsim$ scoring remains expensive, especially for large leaves and long point clouds. In Section~\ref{sec:practical}, we describe how the same batched organization is combined with quantization, compact representations, and specialized low-bit scoring kernels.

Overall, batched search keeps the same retrieval semantics as single-query \mvivf. The advantage is in how the work is organized. By grouping queries by leaf and using vectorized many-to-many $\maxsim$ kernels, \mvivf better exploits cache locality, memory bandwidth, and SIMD parallelism in high-throughput retrieval workloads.


\section{Practical Optimizations}\label{sec:practical}
In \mvivf, multi-vector scoring remains the expensive bottleneck because each query-document comparison involves many vector interactions.
In this section, we describe the main optimizations used in our implementation to accelerate routing, leaf scoring, and batched retrieval.

\subsection{Quantization}

A key bottleneck in multi-vector retrieval is the cost of computing
Chamfer distances, which requires evaluating many pairwise
single-vector distances. We leverage three quantization schemes at
different points on the accuracy--compression tradeoff curve.  All
three share the same preprocessing step: a \emph{randomized rotation}
via a Fast Hadamard Transform (FHT) with random sign flips, following
the approach used in RaBitQ~\cite{gao2024rabitq} and
TurboQuant~\cite{zandieh2025turboquant}. 


\myparagraph{Rand-1bit}
Here we quantize each rotated
coordinate to a single bit by simply taking its sign.
This yields $D'/8$ bytes per vector (i.e., $32\times$ compression over
float32).  
Because the rotation isotropizes the coordinates, sign bits
after random rotation preserve angular relationships between vectors: the expected fraction of differing sign bits
between two vectors is $\arccos(\langle x, y\rangle)/\pi$, so the
Hamming distance of the sign-bit codes is a faithful proxy for angular distance.
Rand-1Bit is
amenable to very fast SIMD implementations as the
distance kernel reduces to XOR plus population count, avoiding all
multiply-add arithmetic.  On hardware supporting VPOPCNTDQ, each
inner-loop iteration processes 512 dimensions via a single XOR and a
single per-64-bit-lane popcount. Furthermore, the per-cloud running
minimum in the Chamfer computation can be maintained entirely in
integer arithmetic, with a single conversion to floating point at the
end.

\myparagraph{TurboQuant 4-bit}
We integrate an existing 4-bit scalar quantization
scheme~\cite{turboquant4bit} that uses a fixed, non-uniform codebook of
16 centroids (8 positive, 8 negative, symmetric about zero) whose
spacing is tailored to the approximately sub-Gaussian distribution of coordinates after rotation.  
Each rotated coordinate is mapped to the nearest centroid and two 4-bit codes are
nibble-packed per byte, giving $D'/2$ bytes per vector.  
Queries are
encoded asymmetrically at 8-bit precision.  
We adapt the single-vector
SIMD kernels from this implementation to the multi-vector setting by tailoring the GEMM kernel to small matrices, circumventing the heavy cache-blocking machinery and unrolling.
We pre-transpose database vectors into panels of sub-vectors to compute 16 4-dimensional dot products per instruction using VPDPBUSD.

\myparagraph{Rand-8bit}
Our 8-bit scheme uses adaptive per-vector uniform quantization:
after rotation, each coordinate is scaled by $$127/\max_i |x_{\mathrm{rot}}[i]|$$
and rounded to the nearest integer in $[-127, 127]$. This is $2\times$
more storage than our 4-bit format, but yields very high-quality
quantization (almost identical to scoring with floats), and also
simplifies the implementation. In particular, since each dimension
occupies a full byte, the SIMD kernel requires no codebook lookup or
nibble unpacking. The multi-vector panel layout and GEMM micro-kernels
are similar to the 4-bit case.

\myparagraph{Other methods}
We also explored other popular quantization strategies, including Product Quantization with FastScan-style optimizations and RaBitQ~\cite{gao2024rabitq}.
We discuss these variants, together with some empirical comparisons, in \supponly{Appendix~\ref{app:quantization}}.

\subsection{Query Compression}\label{sec:query_compression}
Query point clouds often contain substantial redundancy, so many $\maxsim$ computations can be accelerated by scoring a smaller representative set of query vectors. Following this observation, also used in \cite{dhulipala2024muvera}, we compress each query point cloud before routing and leaf scoring.

Given a query $Q$, both compression methods considered below produce a partition $\mathcal{B}=\{B_1,\ldots,B_r\}$ of the query embeddings, and replace each block $B_i$ by a single representative $\bar q_i := \sum_{q\in B_i} q.$ The compressed query is then $\bar Q=\{\bar q_1,\ldots,\bar q_r\}$. Using sums rather than averages preserves the scale of the original $\maxsim$ score: a block with more query embeddings contributes proportionally more to the final score. Larger compression thresholds produce fewer representatives and faster scoring, but can reduce retrieval quality.

\myparagraph{Ball Carving}
Ball carving is the greedy query-compression method used by MUVERA~\cite{dhulipala2024muvera}. Given a distance threshold $\tau$, the algorithm repeatedly selects an unassigned query embedding $q$, forms a block consisting of all currently unassigned embeddings within distance $\tau$ of $q$, replaces this block by the sum of its vectors, and removes the block from the active set. This process continues until all query embeddings have been assigned. Thus, the number of compressed query vectors is not fixed in advance; it is determined by the geometry of the query and the threshold $\tau$.

\myparagraph{Ward's Method}
We also consider compression based on Ward's agglomerative clustering method~\cite{ward1963hierarchical}. Each query embedding is initialized as a singleton cluster. We then repeatedly merge the pair of clusters that causes the smallest increase in within-cluster variance, stopping once the next merge exceeds a threshold $\tau$. As with ball carving, each final cluster is represented by the sum of its embeddings. Ward compression tends to form more balanced and coherent groups of query vectors, and in our experiments gives a slightly better quality--compression tradeoff.

In our implementation, we use Ward's method by default. We also align the compressed query size to the blocking requirements of our many-to-many kernels, avoiding inefficient epilogue cases in the vectorized scoring routines. An empirical comparison of the two compression methods is provided in \supponly{Appendix~\ref{app:query-compression}}.


\section{Experiments}\label{sec:eval}

We evaluate \mvivf through \mvsic, our unified framework for \textbf{M}ulti-\textbf{V}ector \textbf{S}earch, \textbf{I}ndexing, and \textbf{C}lustering (pronounced ``music''). Our evaluation focuses on four questions: (1) how does \mvivf compare with existing multi-vector retrieval methods in single-query latency; (2) how does it perform for high-throughput batched search; (3) what are its indexing time and space requirements; and (4) how do its main design parameters and optimizations affect performance?

\subsection{Experimental Setup}\label{sec:setup}

\myparagraph{Datasets}
\begin{table}[t]
\centering
\footnotesize
\begin{tabular}{cccc}
\toprule
  Dataset &
  Corpus Size &
  Num Queries &
  Avg \# Embs per Doc \\
\midrule
  {\tt nfcorpus}      & 3.6K & 323 & 235.49 \\
     {\tt scifact}    & 5.1K & 300 & 235.65 \\
     {\tt arguana}    & 8.6K & 1,406 & 154.72 \\
     {\tt scidocs}    & 25.6K & 1,000 & 165.05 \\
     {\tt fiqa}       & 57.6K & 648 & 126.9 \\
     {\tt nq}         & 2.68M & 3,452 & 100.3 \\
     {\tt hotpotqa}   & 5.23M & 7,405 & 68.65 \\
     {\tt msmarco}    & 8.84M & 6,980 & 78.8 \\
     {\tt lotte (pooled)}   & 2.4M    & 2,931 & 141.2 \\
\bottomrule
\end{tabular}
\caption{\small Datasets used in this paper.}
\label{tab:datasets}
\end{table} 
We evaluate on eight text retrieval datasets from BEIR~\cite{thakur2021beir} and the pooled LoTTE benchmark from ColBERTv2~\cite{santhanam2022colbertv2}. All documents and queries are encoded using ColBERTv2, yielding one point cloud of token embeddings per document or query. The datasets are summarized in Table~\ref{tab:datasets}.

\myparagraph{Metrics}
Search quality is measured with respect to the exact top-$K$ documents under $\maxsim$. We report Recall-$1$@$K$, the fraction of queries for which the exact nearest neighbor appears in the returned top-$K$ set, and Recall-$K$@$K$, the average fraction of the exact top-$K$ set recovered. While much of the multi-vector retrieval literature emphasizes Recall-$1$@$K$, Recall-$K$@$K$ is the natural analogue of the standard approximate top-$K$ retrieval metric: it measures how well the full top-$K$ neighborhood under $\maxsim$ is preserved.

We restrict the reported operating points to \emph{latency at most $100$ms} and \emph{throughput at least $10$ QPS}, keeping the Pareto curves focused on practically relevant regimes. 

\myparagraph{Methods}
We compare \mvivf against three multi-vector retrieval methods representing different indexing strategies:
\begin{enumerate}[topsep=2pt,itemsep=1pt,parsep=1pt,leftmargin=14pt]
    \item \textbf{SVH} is the single-vector candidate-generation approach underlying PLAID~\cite{santhanam2022plaid} and WARP~\cite{scheerer2025warp}. It indexes the flattened document embeddings using a single-vector ANN index and performs an independent ANN search for each query embedding. Within each search, vectors from the same document are deduplicated by retaining their maximum similarity; these scores are then summed across query embeddings to obtain an approximate $\maxsim$ score for each candidate document. The highest-scoring candidates are then reranked using $\maxsim$ to produce the final top-$k$ results.

    \item \textbf{MUVERA}~\cite{dhulipala2024muvera} maps each query and document point cloud to a fixed-dimensional encoding (FDE) whose inner product approximates their $\maxsim$ score. It indexes the document FDEs using a single-vector ANN index and searches this index using the query FDE to generate candidate documents. The highest-scoring candidates are then reranked using $\maxsim$ to produce the final top-$k$ results.

    \item \textbf{\mvvamana} is our native graph-based baseline. It extends Vamana~\cite{subramanya2019diskann} to point-cloud data by constructing a proximity graph over document point clouds and performing beam search directly using $\maxsim$ scores.
\end{enumerate}

\begin{figure*}[t]
    \centering
    \includegraphics[width=\linewidth]{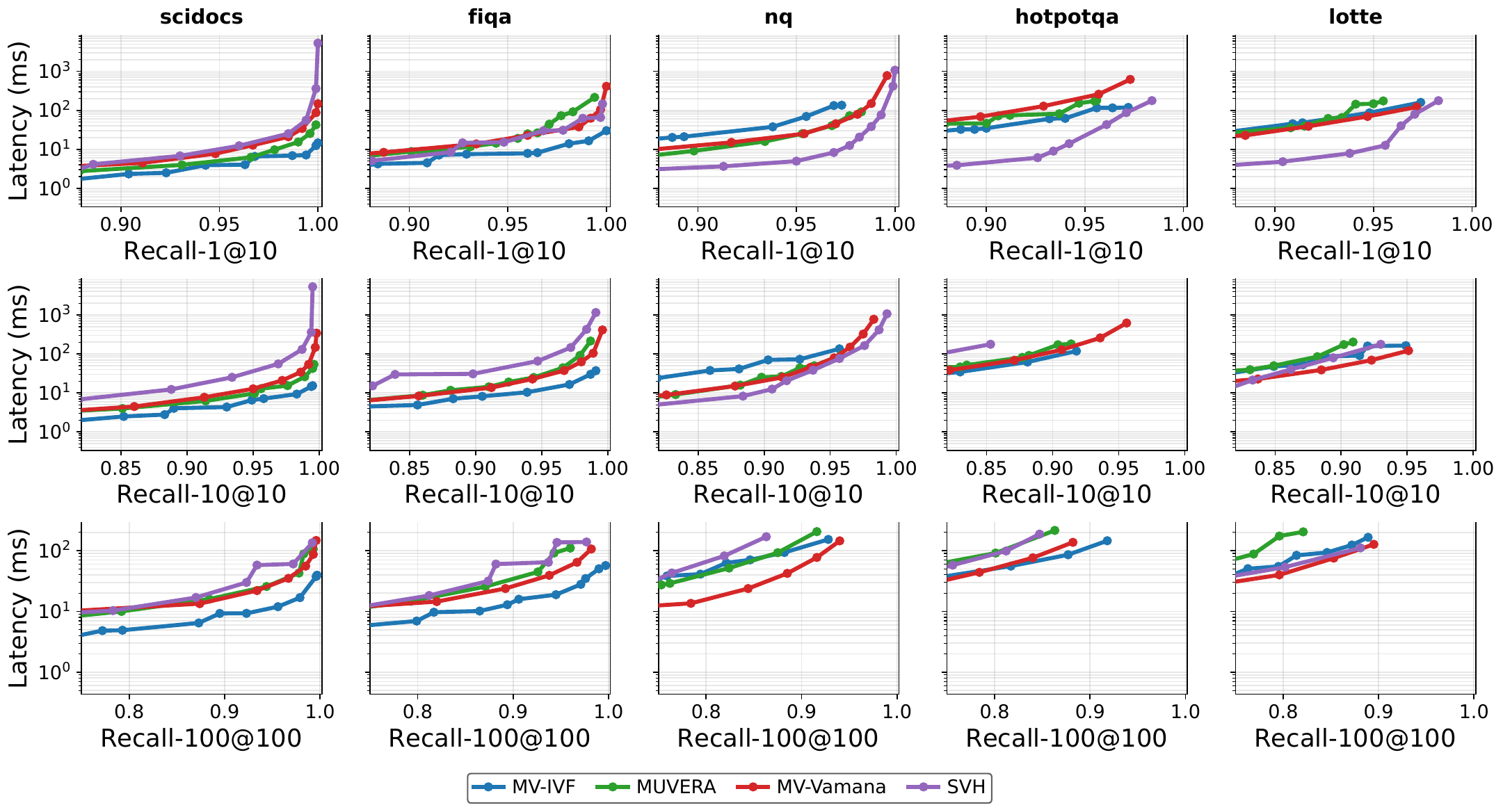}
    \caption{\small Latency vs Recall on representative datasets. Lower-right is better.}
    \label{fig:latency-grid}
    \vspace{-1em}
\end{figure*}

\begin{figure*}[t]
    \centering
    \includegraphics[width=\linewidth]{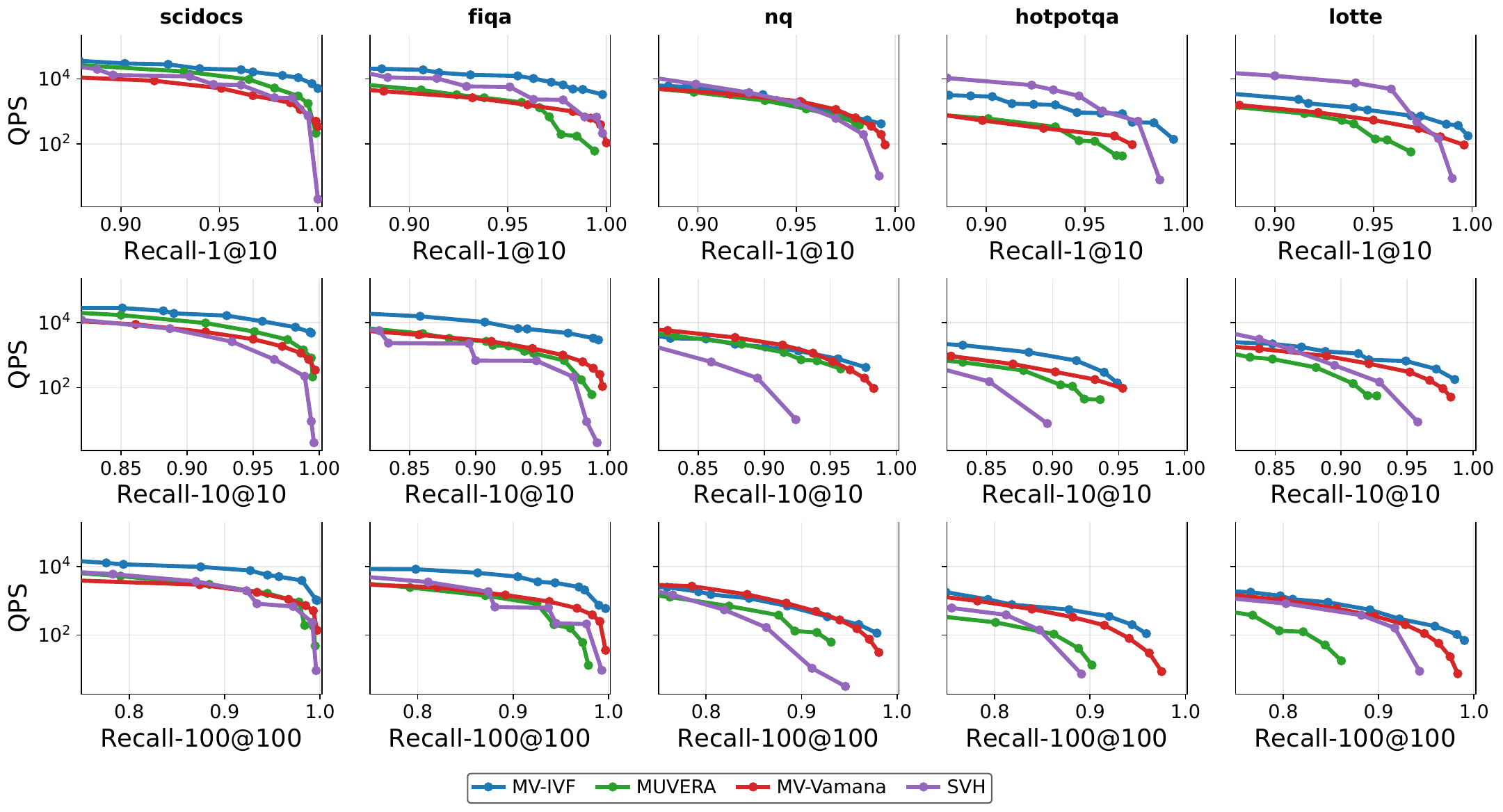}
    
    \vspace{-0.5em}
    \caption{\small Throughput vs. Recall on representative datasets. Upper-right is better.}
    \label{fig:throughput-grid}
    \vspace{-1em}
\end{figure*}

\myparagraph{Implementation and parameter settings}
All four methods are implemented in \mvsic and use the same document/query embeddings, shared optimized scoring infrastructure, and evaluation pipeline. For SVH and MUVERA, the underlying single-vector ANN index is Vamana. For MUVERA, we use $d_{\rm FDE}=10240$, except on \texttt{msmarco}, where we use $d_{\rm FDE}=2560$, following~\cite{dhulipala2024muvera}. For the Vamana indexes, we use standard build parameters from ParlayANN~\cite{parlayann}. The rest of the build and search parameters are described in \supponly{Appendix~\ref{app:baselines}}.

\myparagraph{Quantization}
We use low-bit quantization to accelerate scoring for all methods. For \mvivf, we use TQ-4bit during tree search, Rand-1bit for coarse leaf scoring, and Rand-8bit for reranking. For SVH, MUVERA, and \mvvamana, we use TQ-4bit during Vamana search and Rand-8bit for reranking. The quality and performance of the quantization schemes used in our implementation are evaluated in \supponly{Appendix~\ref{app:quantization}}.

\myparagraph{Machine Information}
We ran all experiments other than on the MS MARCO dataset on a dual-socket Intel® Xeon® Platinum 8481C CPU @ 2.70 GHz machine with 88 physical cores (176 hyperthreads) and 1.4 TiB of main memory. The machine has 6.9 MiB of L1 cache, 176 MiB of L2 cache, and 210 MiB of L3 cache.

We ran the experiments for {\tt msmarco} (our largest dataset) on a dual-socket Intel Xeon 6985P-C CPU @ 2.30GHz machine with 144 physical cores (288 hyperthreads), 2.1 TiB of main memory, and 15.8 MiB of L1 cache, 288 MiB of L2 cache, and 1008 MiB of L3 cache.

\subsection{End-to-End Search Performance}\label{sec:end-to-end}

\myparagraph{Latency Experiments}
We first evaluate query latency. For each method and dataset, we sweep the method's search parameters and plot the Pareto frontier of latency versus Recall-$1$@$10$, Recall-$10$@$10$, and Recall-$100$@$100$. The system uses only a single core here. Figure~\ref{fig:latency-grid} shows these curves on a representative subset of the datasets; see \supponly{Appendix~\ref{app:latency}} for full results.

The three recall metrics highlight different aspects of the retrieval problem. For Recall-$1$@$10$, SVH is a particularly strong baseline. This is expected: SVH performs candidate generation at the level of individual embedding vectors, and is therefore well suited to finding documents with a small number of highly matching token-level interactions. After reranking, this is often enough to recover the nearest document under $\maxsim$.

The picture changes for Recall-$10$@$10$. On most datasets, SVH is no longer competitive, indicating that recovering the full top-$10$ neighborhood requires more than finding a few high-signal token matches. In this regime, \mvivf, MV-Vamana, and MUVERA are all competitive on the larger datasets ({\tt hotpotqa, lotte}), while \mvivf outperforms all methods on the smaller datasets ({\tt nfcorpus, scifact, arguana, scidocs, fiqa}). The main exception is {\tt nq}, where SVH remains strong even for Recall-$10$@$10$. This suggests that on this dataset, the top-$10$ documents are often recoverable from a small number of strong token-level matches, making single-vector candidate generation unusually effective.

Recall-$100$@$100$ gives the most demanding test of approximate multi-vector retrieval. Here, the goal is to preserve a much larger neighborhood under $\maxsim$. In this regime, SVH struggles even on {\tt nq}, consistent with the view that token-level candidate generation is most effective for recovering the very top of the ranking, but less effective for preserving a broader neighborhood under $\maxsim$. On larger datasets, MV-Vamana often gives the best single-query latency frontier, consistent with the advantages of graph-based indexes in minimizing the number of point-cloud score evaluations. 
At the same time, \mvivf is competitive with the best method across datasets for Recall-$100$@$100$, showing that the clustering-based partition preserves enough multi-vector structure for high-quality routing.

We note that SVH remains strong on {\tt msmarco} even for Recall-$100$@$100$. This may partly reflect that {\tt msmarco} is part of the training data for ColBERTv2, making token-level matching unusually effective on this dataset.

Overall, at $90\%$ Recall-$10$@$10$, \mvivf is $1.33\times$ faster than MUVERA, $1.55\times$ faster than SVH, and $1.22\times$ faster than MV-Vamana in geometric mean, with maximum speedups of up to $4.76\times$, $8.03\times$, and $5.13\times$, respectively. 

\begin{figure}[t]
    \centering
    \includegraphics[width=\linewidth]
        {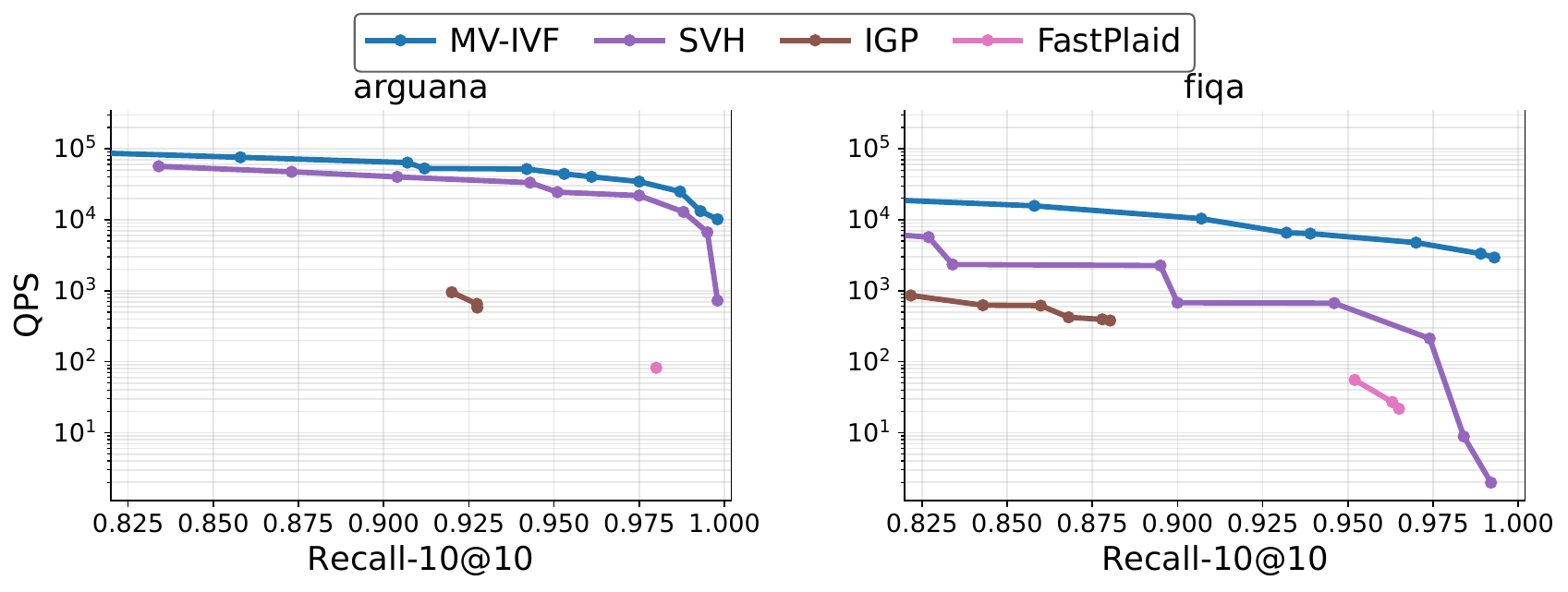}
    \caption{\small Batched throughput versus Recall-$10$@$10$ for
    \mvivf, SVH, FastPLAID, and IGP on \texttt{arguana} and
    \texttt{fiqa}. Upper-right is better.}
    \label{fig:plaid-igp}
    \vspace{-0.5em}
\end{figure}

\myparagraph{Throughput Experiments}

We next evaluate batched throughput. In this setting, the input is a batch of queries and the system uses all available cores; we report queries per second (QPS) as a function of Recall-$1$@$10$, Recall-$10$@$10$, and Recall-$100$@$100$.
For the baselines, batching essentially parallelizes independent query executions. For \mvivf, batching applies the cache-efficient batch search procedure described in Section~\ref{sec:mvivf}. Figure~\ref{fig:throughput-grid} shows results on a representative subset of datasets; see \supponly{Appendix~\ref{app:throughput}} for full results.

In the batched setting, \mvivf consistently gives the strongest throughput frontier across datasets and recall metrics. The improvement is especially pronounced on the smaller datasets, where \mvivf dominates the other methods across nearly the full recall range. Two aspects of \mvivf help explain this behavior. First, the IVF structure is highly parallelizable: once the probe leaves have been identified, the leaf scans expose a large amount of independent scoring work. Second, the batched search procedure of Section~\ref{sec:mvivf} improves locality by grouping queries that visit the same leaf. This effect is particularly important in the high-recall regime, where each query probes many leaves and the same leaf data is more likely to be reused across queries.

\begin{figure*}[t]
    \vspace{-0.5em}
    \centering
    \begin{subfigure}[t]{0.49\linewidth}
        \centering\includegraphics[width=\linewidth]{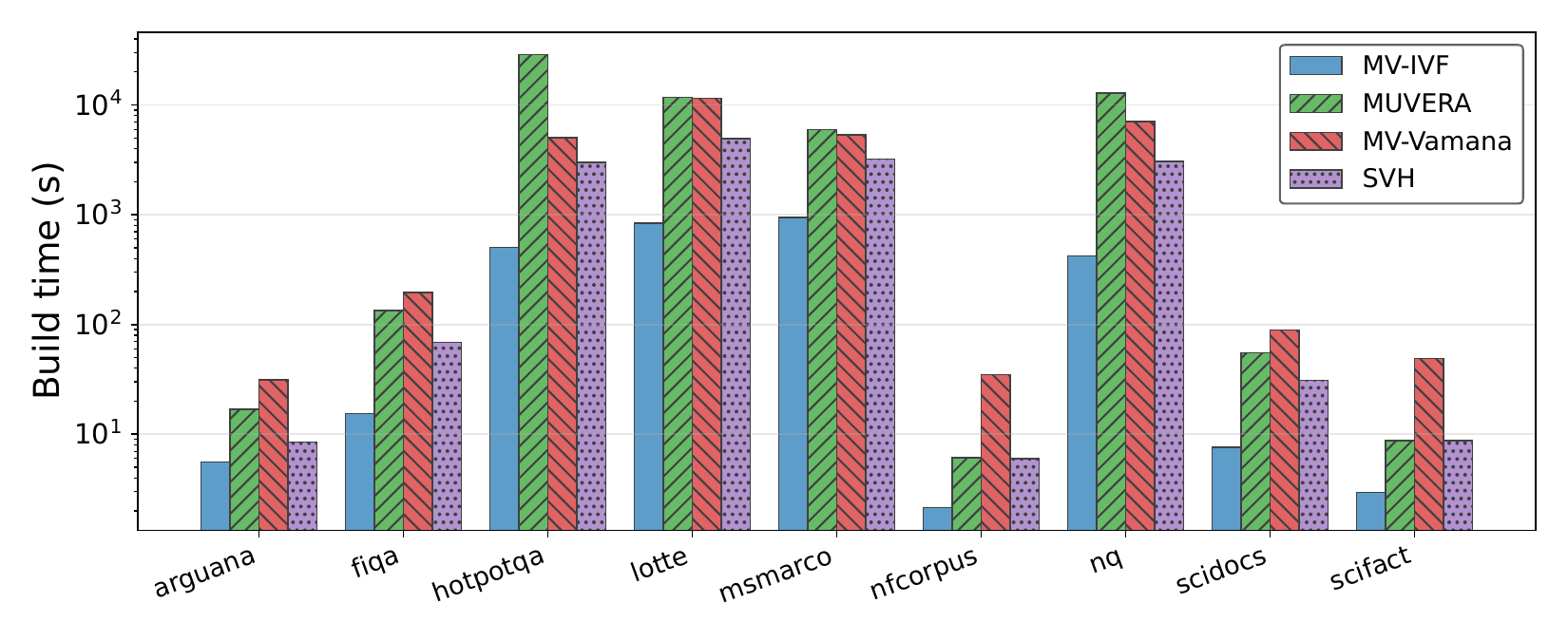}
    \end{subfigure}
    \hfill
    \begin{subfigure}[t]{0.49\linewidth}
        \centering\includegraphics[width=\linewidth]{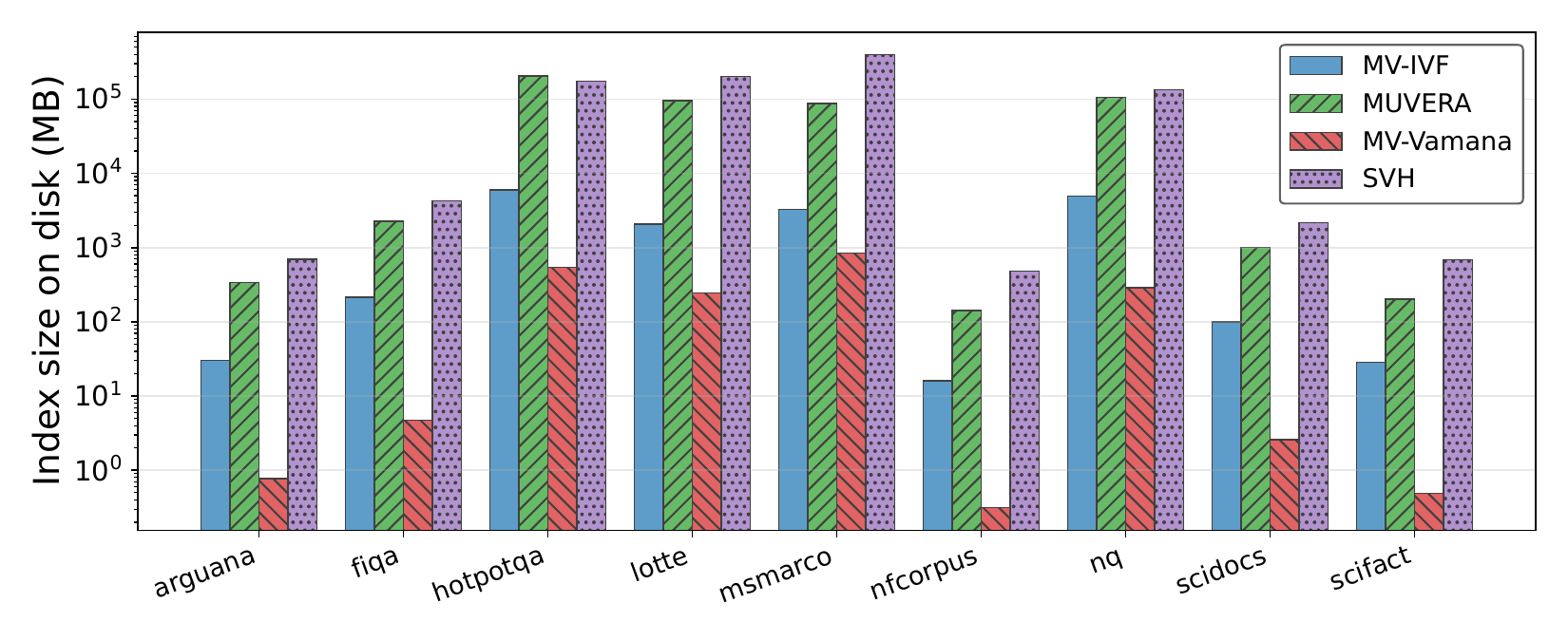}
    \end{subfigure}
    \vspace{-0.6em}
    \caption{\small Wall-clock build time (left) and on-disk index size (right).}
    \label{fig:build-cost}
    \vspace{-1em}
\end{figure*}

\begin{figure}[t]
    \centering
    \includegraphics[width=\linewidth]
        {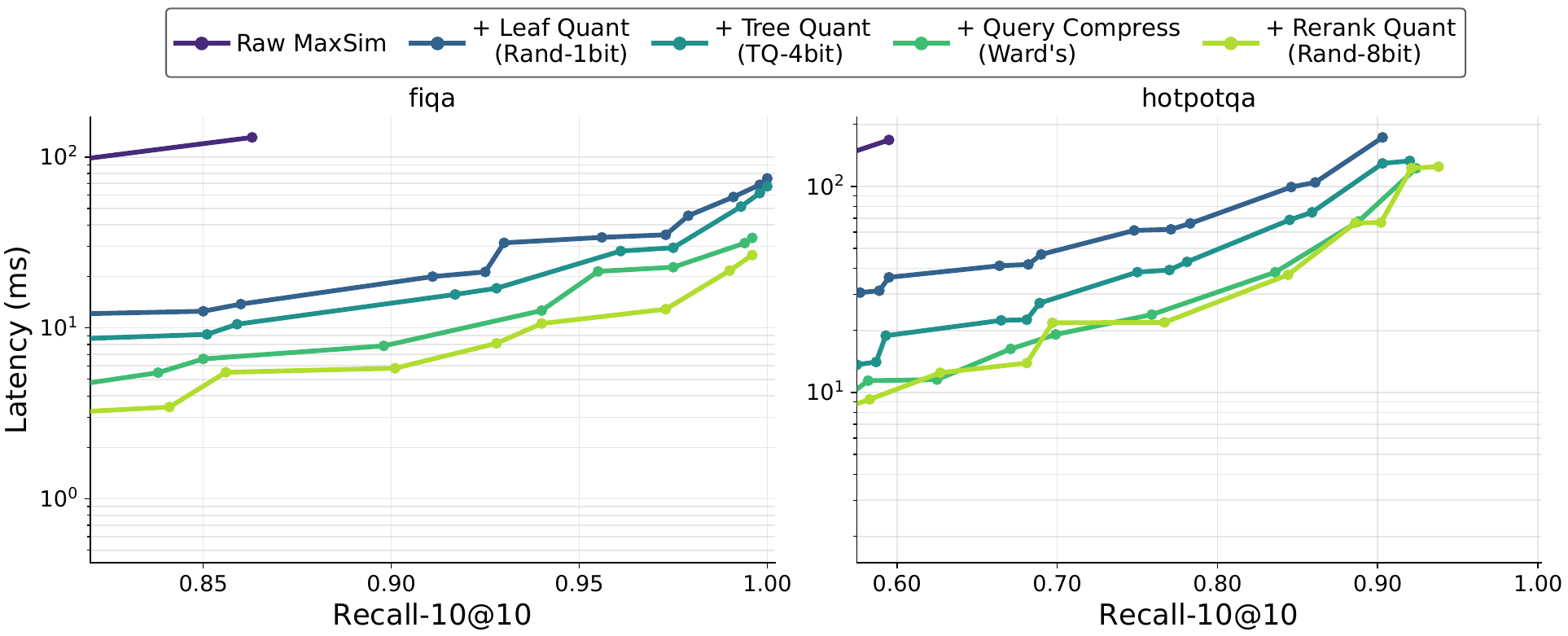}
    \caption{\small Effect of the \mvivf optimizations on single-query
    latency versus Recall-$10$@$10$ on \texttt{fiqa} and
    \texttt{hotpotqa}. Each curve cumulatively adds the indicated
    optimization. Lower-right is better.}
    \label{fig:mvivf-optimization-ablation}
    \vspace{-0.5em}
\end{figure}

MV-Vamana is generally the closest competitor, but its throughput degrades in the very high-recall regime. This is consistent with the behavior of graph-based search: while it is effective at reducing the number of expensive multi-vector comparisons for a single query, its data-dependent traversal is less amenable to batching and leads to less regular memory access. SVH and MUVERA are less competitive in the batched setting, especially on the larger datasets and for Recall-$100$@$100$, where recovering a broad neighborhood under $\maxsim$ requires substantially more candidate-generation and reranking work.

At $90\%$ Recall-$10$@$10$, \mvivf improves QPS by $2.19\times$ over MUVERA, $4.01\times$ over SVH, and $2.07\times$ over MV-Vamana in geometric mean, with best observed improvements of up to $8.40\times$, $173.5\times$, and $4.99\times$, respectively. 

\myparagraph{Comparison with PLAID and IGP}

We also compare against PLAID~\cite{santhanam2022plaid} and IGP~\cite{bian2025igp}. For PLAID, we use FastPLAID~\cite{fastplaid2025}, a highly optimized implementation of PLAID; for IGP, we use the implementation released by the authors~\cite{bian2025igp}. These implementations did not scale reliably across the full collection of datasets and low-latency operating points considered in our main experiments, so we report representative comparisons on \texttt{arguana} and \texttt{fiqa}.

Figure~\ref{fig:plaid-igp} shows that \mvivf substantially outperforms both external implementations on both datasets. Over the recall ranges where they overlap, \mvivf achieves tens of times higher throughput than IGP and over two orders of magnitude higher throughput than FastPLAID. SVH also substantially outperforms FastPLAID, supporting its use as an optimized implementation of the single-vector candidate-generation approach underlying PLAID-style methods.

\subsection{Build Time and Index Size}\label{sec:build-cost}

We next compare the preprocessing cost and storage overhead of each method. Figure~\ref{fig:build-cost} reports wall-clock build time and final on-disk index size across datasets.

Overall, \mvivf has a favorable build-time and index-storage profile. Its construction is practical across datasets and is often substantially faster than methods that use single-vector retrieval machinery as a black box. The IVF tree introduces little additional storage beyond the center point clouds and leaf-level membership information. \mvvamana produces the smallest index, but its construction is substantially more expensive because building the graph requires many multi-vector comparisons. Thus, \mvivf provides a useful middle ground: it retains native multi-vector routing while keeping both preprocessing time and index size moderate.

\subsection{Ablation Studies}\label{sec:ablations}

We next study how the main design choices and systems optimizations in
\mvivf affect retrieval performance. We report these experiments on
\texttt{fiqa} and \texttt{hotpotqa}, representing a smaller and a larger
retrieval dataset, respectively.

\myparagraph{Impact of optimizations}
Figure~\ref{fig:mvivf-optimization-ablation} shows the latency--recall
frontier as we cumulatively add the optimizations described in
Section~\ref{sec:practical}. Starting from full-precision $\maxsim$ scoring,
we successively add Rand-1bit leaf scoring, TQ-4bit tree scoring, Ward query
compression, and Rand-8bit reranking.

\begin{figure}[t]
    \centering
    \includegraphics[width=\linewidth]
        {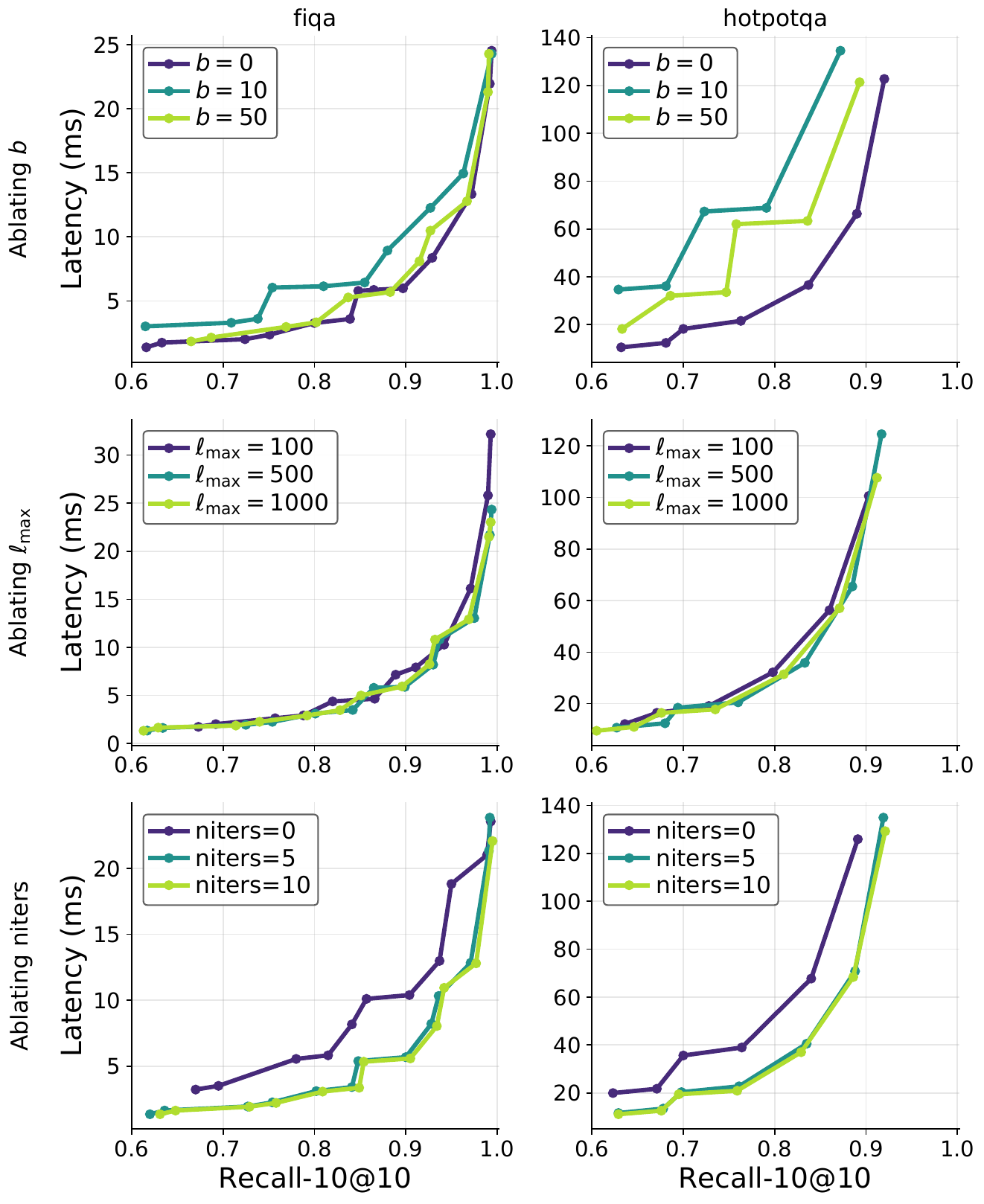}
    \caption{\small Sensitivity of \mvivf to the branching parameter $b$
    (top), maximum leaf size $\ell_{\max}$ (middle), and number of
    \mvlloyd iterations (down), on \texttt{fiqa} (left) and
    \texttt{hotpotqa} (right). Lower-right is better.}
    \label{fig:mvivf-parameter-ablation}
    \vspace{-0.5em}
\end{figure}

The optimizations substantially improve the latency--recall frontier. In particular, the unoptimized implementation does not reach $90\%$ Recall-$10$@$10$ within our $100$ms latency budget on either dataset. At $90\%$ Recall-$10$@$10$ on \texttt{fiqa}, Rand-1bit leaf scoring alone provides at least a $5.37\times$ speedup over the unoptimized implementation. Adding tree quantization, query compression, and quantized reranking gives further speedups of $1.31\times$, $1.76\times$, and $1.39\times$, respectively. We observe similar incremental improvements on \texttt{hotpotqa}. Overall, the complete optimization stack is at least $17.2\times$ faster on \texttt{fiqa} at $90\%$ recall.

\myparagraph{Sensitivity to MV-IVF parameters}
We next study the three main parameters controlling the structure and
construction of the \mvivf index: the branching parameter $b$, the maximum
leaf size $\ell_{\max}$, and the number of \mvlloyd iterations. We vary one
parameter at a time while holding the remaining parameters fixed at their
default values.

Figure~\ref{fig:mvivf-parameter-ablation} shows that the default settings are
robust across the two datasets. Setting $b=0$ enables our automatic branching
rule, which chooses the branching factor based on the number of point clouds
at each node (roughly its square root, subject to the target leaf size).
Automatic branching is competitive with fixed branching on \texttt{fiqa} and
gives the strongest frontier on \texttt{hotpotqa}. Performance is relatively
insensitive to $\ell_{\max}$ over the range tested; however, smaller leaves
increase both index size and construction time. We therefore use
$\ell_{\max}=500$ as a reasonable default, although $\ell_{\max}=1000$ gives
similar search performance.

Finally, $\texttt{niters}=0$ performs no \mvlloyd updates and simply samples
the initial center point clouds from the input. Running \mvlloyd substantially
improves the latency--recall frontier, showing that refining the center point
clouds is important for effective routing. The improvement from $5$ to $10$
iterations is small, so we use $5$ iterations by default.

\section{Conclusion}\label{sec:conclusion}

We presented \mvivf, a native IVF-style index for multi-vector retrieval. The key ingredient is \mvlloyd, a Lloyd-style clustering primitive for point clouds under a Chamfer-based objective, which lets us build an IVF tree whose routing representatives are themselves point clouds. Together with quantized scoring, query compression, and cache-efficient batched search, \mvivf gives a practical clustering-based approach to multi-vector search. We also presented MV-Vamana, a native graph-based index, and MVSIC, a unified C++ framework for implementing and evaluating multi-vector retrieval methods. Our experiments show that \mvivf provides strong latency--recall and throughput--recall tradeoffs while maintaining moderate build time and index size.

Several directions remain open. First, the clustering problem underlying \mvlloyd deserves further study, including approximation algorithms, initialization schemes, and variants based on different normalizations or set-to-set similarities. Although clustering with non-metric objectives is generally hard, it is not clear whether existing hardness results fully capture the structure of Chamfer objectives arising from multi-vector embeddings. Second, the design space for multi-vector IVF is still far from exhausted: spilling, probe allocation, and reranking strategies may lead to further improvements, especially on larger datasets. Finally, MV-Vamana is a promising direction in its own right. While its build time is high and its irregular search pattern is less favorable for batching, graph-based search is effective at minimizing expensive multi-vector comparisons, and merits deeper study.


\begin{acks}
We thank Vincent Cohen-Addad and Anupam Gupta for helpful discussions.
Kishen N Gowda and Laxman Dhulipala were supported in part by NSF awards CCF-2403235 and CNS-2317194.
\end{acks}


\bibliographystyle{ACM-Reference-Format}
\bibliography{references}

\appendix
\newcommand{\appendixplot}[3]{%
    \IfFileExists{plots/#1}{%
        \begin{subfigure}[t]{#2}
            \centering
            \includegraphics[width=\linewidth]{plots/#1}
            \caption{\texttt{#3}}
        \end{subfigure}
    }{}%
}

\newcommand{\appendixevaluationplot}[4]{%
    \IfFileExists{plots/#1}{%
        \begin{figure*}[!htbp]
            \centering
            \includegraphics[width=\linewidth]{plots/#1}
            \caption{\small #2: \texttt{#3}.}
            \label{#4}
        \end{figure*}
    }{}%
}

\section{Proof of Proposition~\ref{prp:mvlloyd}}\label{app:prop1_proof}
\begin{proof}[Proof of Proposition~\ref{prp:mvlloyd}]
By definition of Chamfer distance, for any center-set $C$,
\begin{align*}
    \sum_{P\in \Pi_j} \Chn(P,C)
    =
    \sum_{P\in \Pi_j} \frac{1}{|P|}\sum_{x\in P}\min_{c\in C} d(x,c).
\end{align*}
Reindexing the double sum over the multiset
$\flatten(\Pi_j)$, the point $x\in P$ receives weight $1/|P|$. Hence,
\begin{align*}
    \sum_{P\in \Pi_j} \Chn(P,C)
    =
    \sum_{x\in \flatten(\Pi_j)} w(x)\min_{c\in C} d(x,c),
\end{align*}
where \(w(x)=1/|P|\) for $x$ coming from $P$. Therefore, minimizing the Chamfer $(1,s)$-median objective over all center-sets $C$ with $|C|\le s$ is exactly the weighted  $s$-median problem on the flattened multiset $\flatten(\Pi_j)$.
\end{proof}


\section{Experiments (continued)}\label{app:experiments}
In this section, we continue the discussion of the experiments from the main paper.

\subsection{Baselines and Parameter Settings}\label{app:baselines}
\begin{table*}[t]
\centering
\footnotesize
\setlength{\tabcolsep}{4pt}
\renewcommand{\arraystretch}{1.12}

\newcommand{\paramitems}[1]{%
\begin{minipage}[t]{\linewidth}
\vspace{0pt}
\begin{itemize}[leftmargin=1.1em,itemsep=1pt,topsep=0pt,parsep=0pt]
#1
\end{itemize}
\end{minipage}
}

\begin{tabular}{
>{\raggedright\arraybackslash}p{0.12\linewidth}
>{\raggedright\arraybackslash}p{0.39\linewidth}
>{\raggedright\arraybackslash}p{0.41\linewidth}}
\toprule
Method & Build parameters & Search parameters \\
\midrule

\mvivf
&
\paramitems{
    \item Recursive \mvlloyd tree with automatic branching.
    \item If a node contains $n_u$ point clouds and $\branch=0$, then
    $\branch_u=\min\{\lceil\sqrt{n_u}\rceil,\lceil c n_u/\leafsize\rceil\}$.
    \item $5$ Lloyd iterations; at most $100$ point clouds per update cluster.
    \item Build-time assignments use TQ-8bit scoring.
    \item $\leafsize=500$.
}
&
\paramitems{
    \item $\nprobe \in \{1,2,4,\ldots,\}$.
    \item TQ-4bit for greedy search; Rand-1bit for leaf scoring; Rand-8bit for reranking.
    \item Query compression: Ward's method with $\tau=0.5$.
    \item \texttt{num\_rerank} $\in \{20,100,500\}$ for $K=10$.
    \item \texttt{num\_rerank} $\in \{200,1000,5000\}$ for $K=100$.
}
\\

\midrule
MUVERA
&
\paramitems{
    \item $d_{\rm FDE}=10240$, except MS MARCO where $d_{\rm FDE}=2560$.
    \item Vamana graph with $R=200$, $L_{\rm build}=600$, and $\alpha=1.0$.
}
&
\paramitems{
    \item Search beam $L \in \{16,32,\ldots,\}$.
    \item TQ-4bit for graph search; Rand-8bit for reranking.
    \item \texttt{num\_rerank} $\in \{500,1000\}$ for $K=10$.
    \item \texttt{num\_rerank} $\in \{5000,10000\}$ for $K=100$.
}
\\

\midrule
SVH
&
\paramitems{
    \item Vamana graph over flattened individual embeddings.
    \item $R=64$, $L_{\rm build}=128$, and $\alpha=1.0$.
}
&
\paramitems{
    \item Search beam $L \in \{16,32,\ldots,\}$.
    \item TQ-4bit for graph search; Rand-8bit for reranking.
    \item \texttt{num\_rerank} $\in \{50,100, 500\}$ for $K=10$.
    \item \texttt{num\_rerank} $\in \{1000, 5000\}$ for $K=100$.
}
\\

\midrule
\mvvamana
&
\paramitems{
    \item Native Vamana graph over document point clouds.
    \item $R=64$, $L_{\rm build}=128$, and $\alpha=1.0$.
}
&
\paramitems{
    \item Search beam $L \in \{16,32,\ldots,\}$.
    \item TQ-4bit for graph search; Rand-8bit for reranking.
    \item \texttt{num\_rerank} $\in \{20,50\}$ for $K=10$.
    \item \texttt{num\_rerank} $\in \{200,500\}$ for $K=100$.
}
\\

\bottomrule
\end{tabular}

\caption{\small Build parameters and
search-time grids used to generate the Pareto curves in
Sections~\ref{sec:eval}. Early cut-off was applied when latency went above 100ms, or QPS dropped below 10.}
\label{tab:methods}
\end{table*}
We compare \mvivf against three methods: SVH, MUVERA, and \mvvamana. SVH is a
single-vector candidate-generation baseline: it indexes the flattened document
embeddings with a single-vector ANN index, retrieves candidate documents through
their individual embeddings, and reranks the resulting candidates using
multi-vector scoring. MUVERA reduces each point cloud to a fixed-dimensional
encoding and performs candidate generation using a single-vector ANN index over
these encodings. \mvvamana is our native graph-based baseline, obtained by
building a Vamana-style graph directly over document point clouds.

All four methods are implemented in \mvsic and use the same document/query
embeddings, scoring kernels, and evaluation pipeline. For SVH and MUVERA, the
underlying single-vector ANN index is Vamana. For MUVERA, we use the parameter
settings recommended in~\cite{dhulipala2024muvera}; for all Vamana graphs, we
use standard settings from ParlayANN~\cite{parlayann}. For each method, we sweep
the search-time parameters listed in Table~\ref{tab:methods} and report the
Pareto frontier over all resulting operating points. Thus, the main comparisons
are based on the best recall--efficiency tradeoff achieved by each method over
its specified parameter grid, rather than on a single hand-tuned setting.

\subsection{Quantization}\label{app:quantization}
We evaluate several quantization paradigms used throughout the paper. In addition to the TQ, Rand-1bit, and Rand-8bit variants described in Section~\ref{sec:practical}, we also compare against PQ FastScan-style kernels~\cite{andre:hal-01239055,andre2017accelerated} and RaBitQ~\cite{gao2024rabitq}. For RaBitQ, we use the publicly available RaBitQ library implementation~\cite{gao2024rabitq}.

\begin{figure*}[t]
    \centering
    \appendixplot{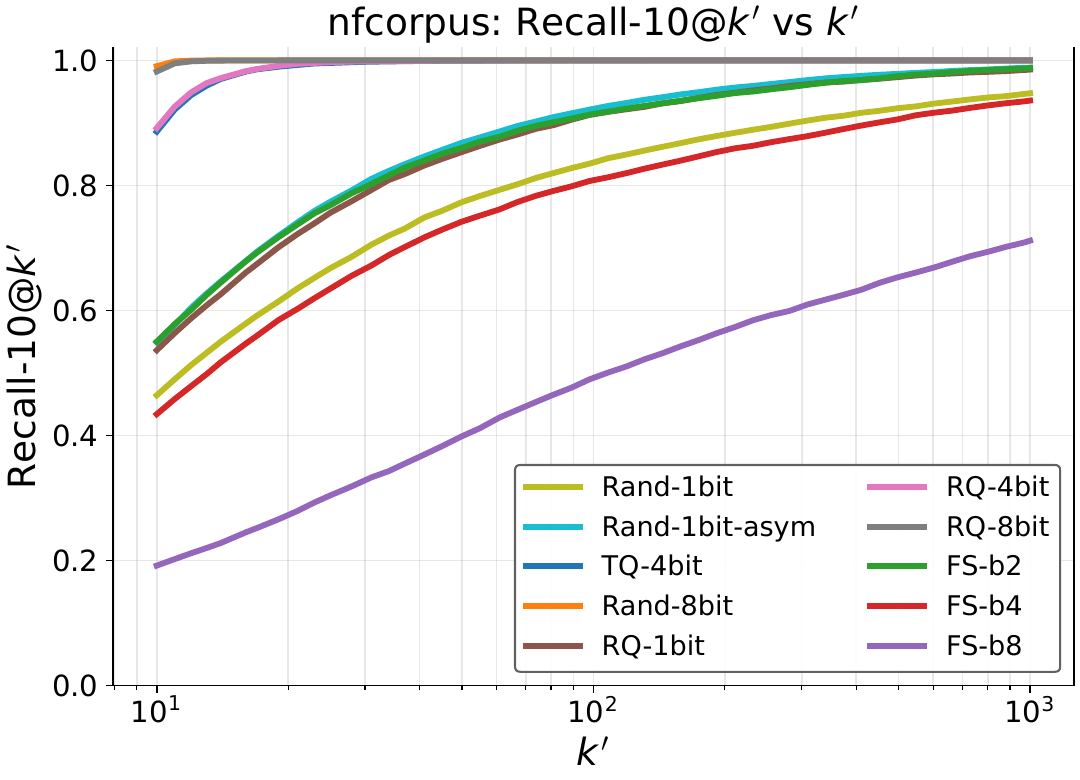}{0.235\linewidth}{nfcorpus}
    \hfill
    \appendixplot{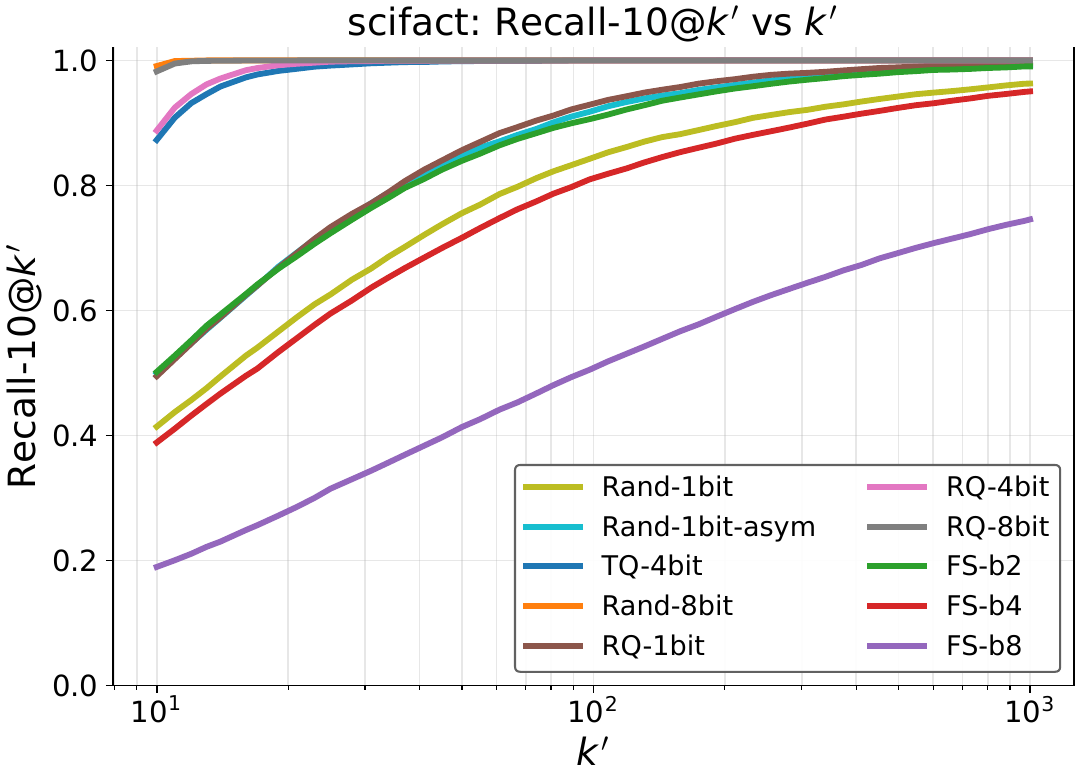}{0.235\linewidth}{scifact}
    \hfill
    \appendixplot{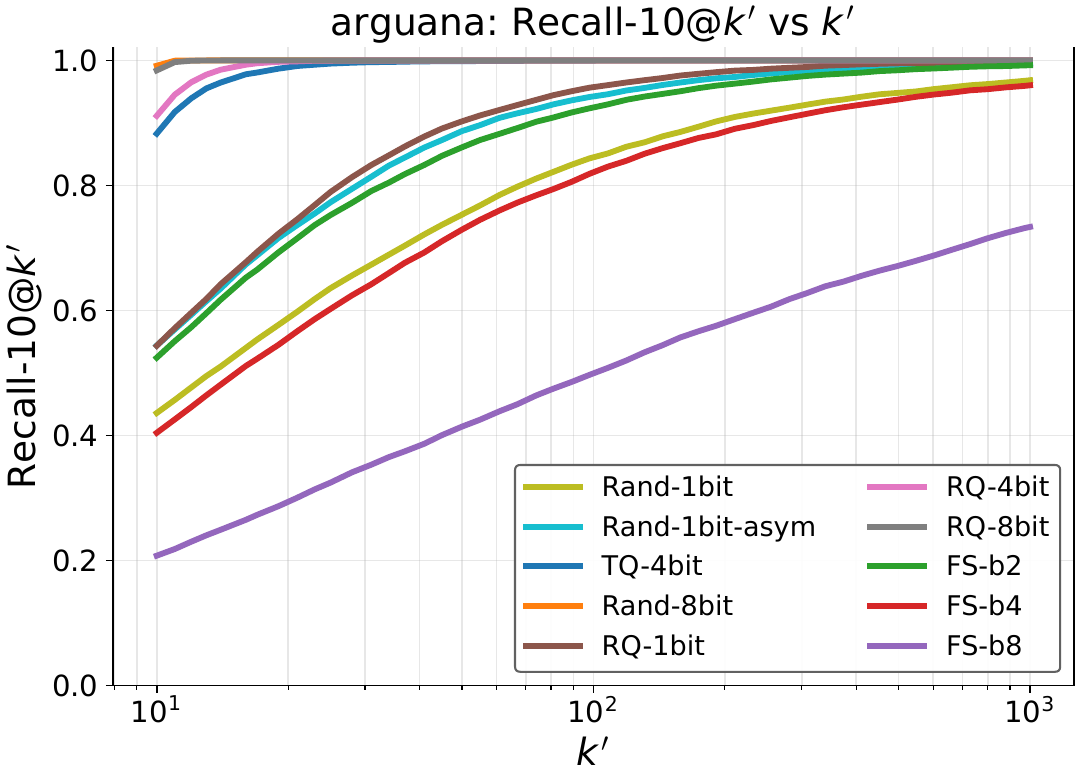}{0.235\linewidth}{arguana}
    \hfill
    \appendixplot{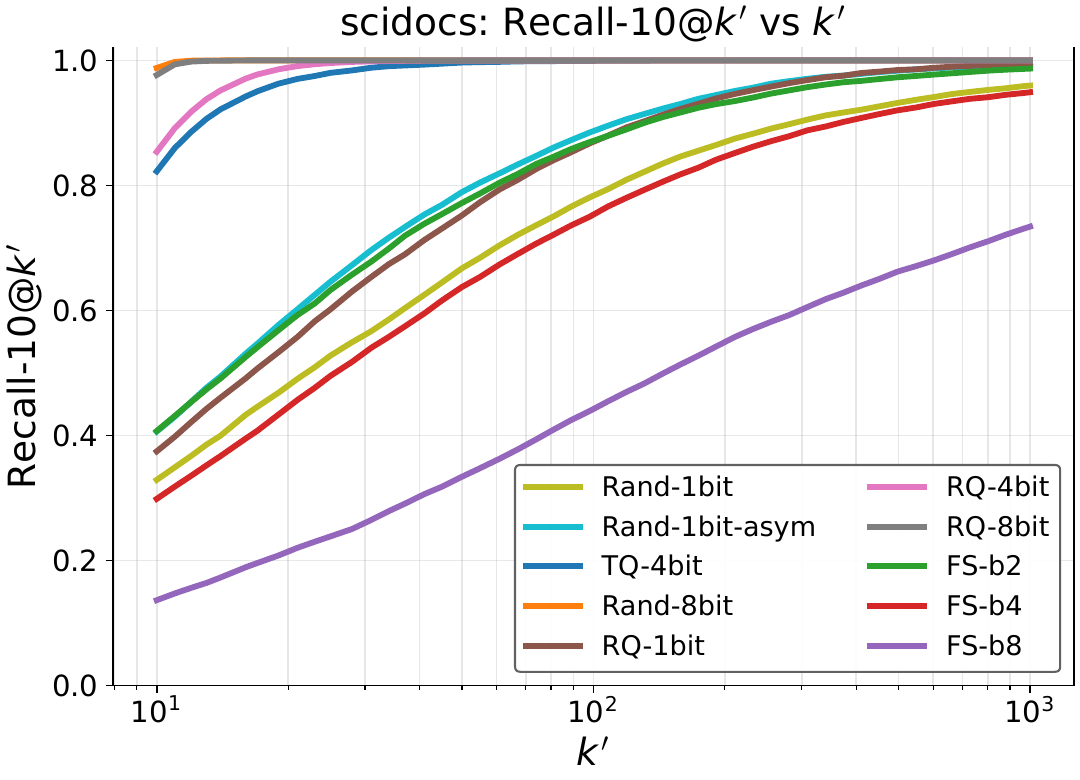}{0.235\linewidth}{scidocs}

    \vspace{0.75em}
    \appendixplot{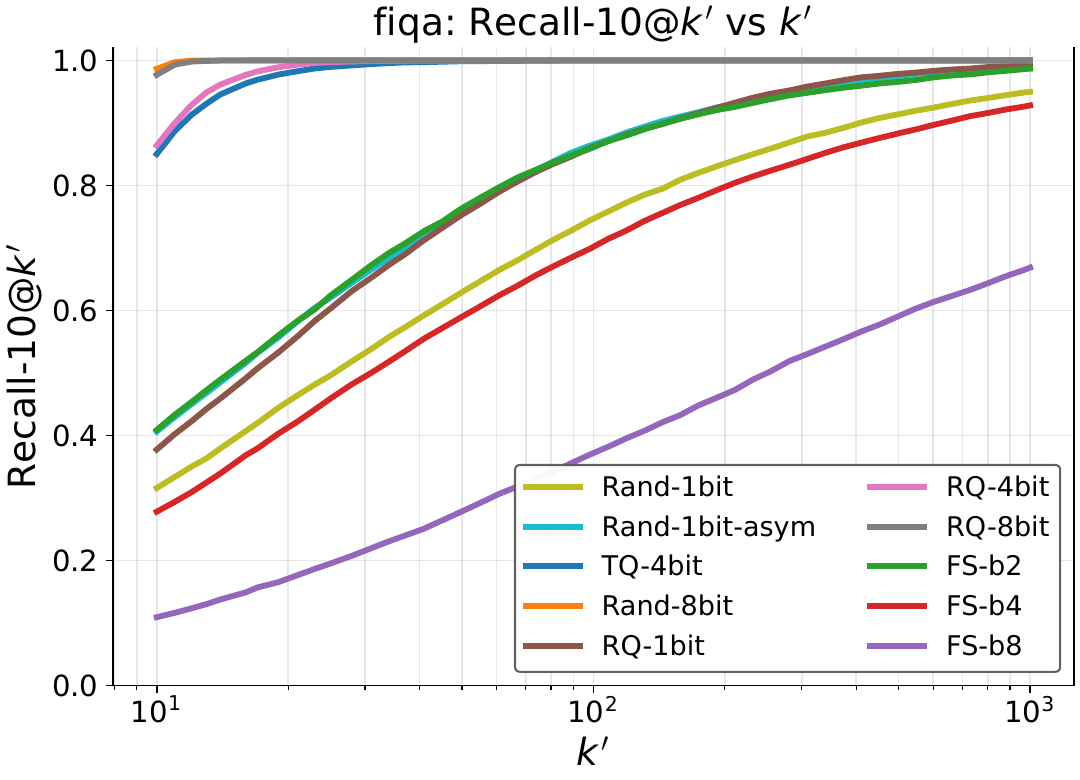}{0.30\linewidth}{fiqa}
    \hfill
    \appendixplot{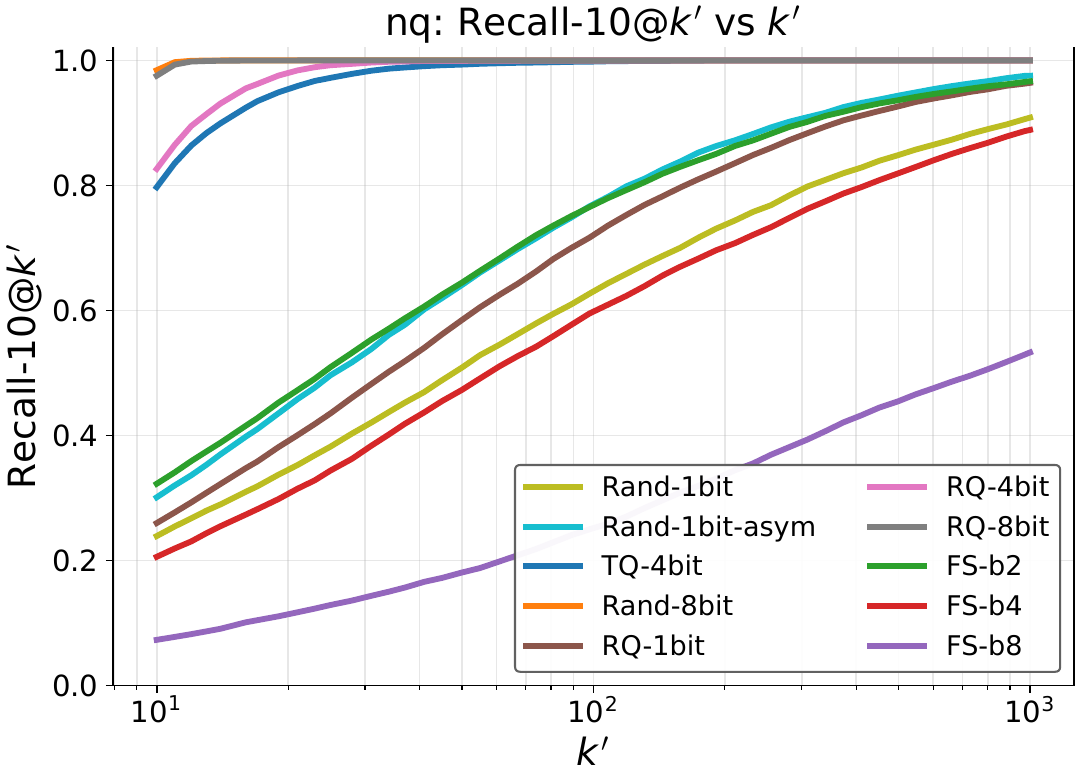}{0.30\linewidth}{nq}
    \hfill
    \appendixplot{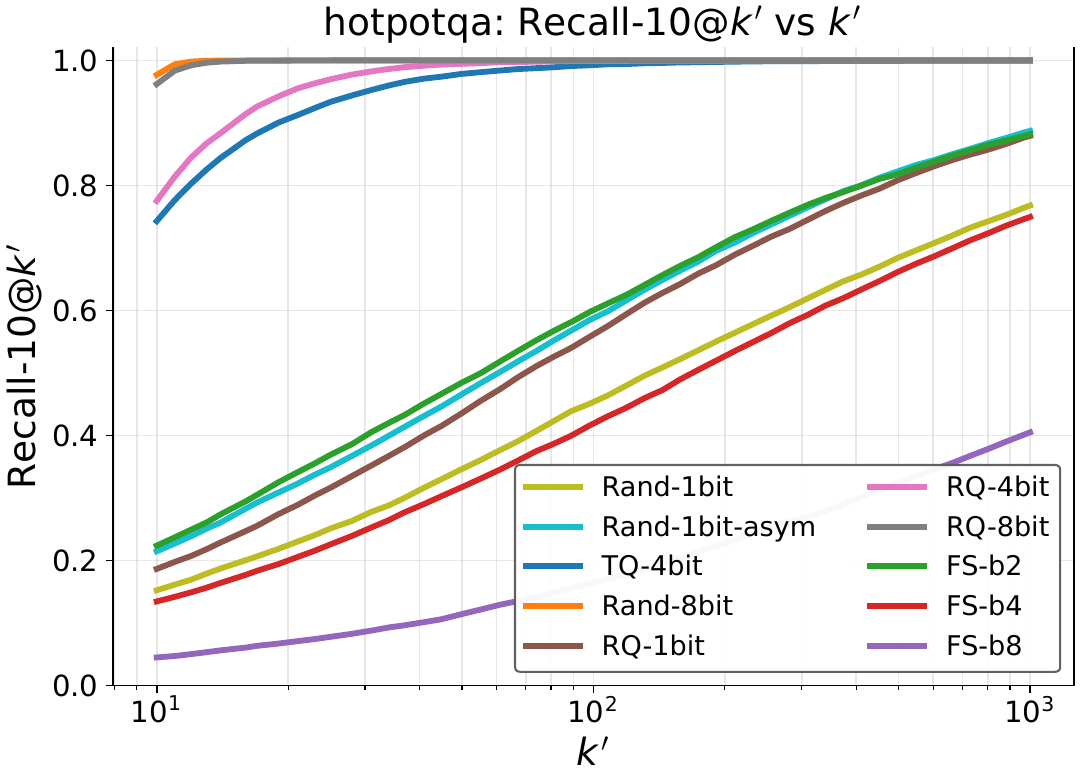}{0.30\linewidth}{hotpotqa}
    \caption{\small Single-vector over-retrieval quality across datasets. Each panel shows recall as a function of the candidate budget $k'$.}
    \label{fig:app-single-vector-overretrieve}
\end{figure*}

\myparagraph{Single-vector over-retrieval benchmark}
We first evaluate the quantizers on single-vector inner-product search. For each
quantized scorer, we retrieve the top $k'$ candidates under the approximate
scores and report Recall-$10$@$k'$ with respect to the exact top-$10$ neighbors.
This measures how much over-retrieval is needed for the quantized scores to
preserve the true nearest neighbors. Figure~\ref{fig:app-single-vector-overretrieve}
reports the resulting recall curves.

In the plot labels, FS-$4b$, FS-$8b$, etc. denote the block-size variants used
by our FastScan implementation; in this notation, FS-$4b$ corresponds to one bit
per coordinate, while FS-$8b$ corresponds to one half-bit per coordinate. We
also include both Rand-1bit and Rand-1bit-asym: Rand-1bit quantizes both
database and query vectors, while Rand-1bit-asym uses an asymmetric query
variant. We observe that Rand-8bit and RQ-8bit preserve quality extremely well,
making them suitable as high-fidelity reranking quantizers. RQ-4bit and TQ-4bit
provide a good medium-quality tradeoff and are well suited for candidate
generation, while Rand-1bit, RQ-1bit, FS-$2b$, and FS-$4b$ are lower-quality
but substantially cheaper scorers.

\begin{figure*}[t]
    \centering
    \appendixplot{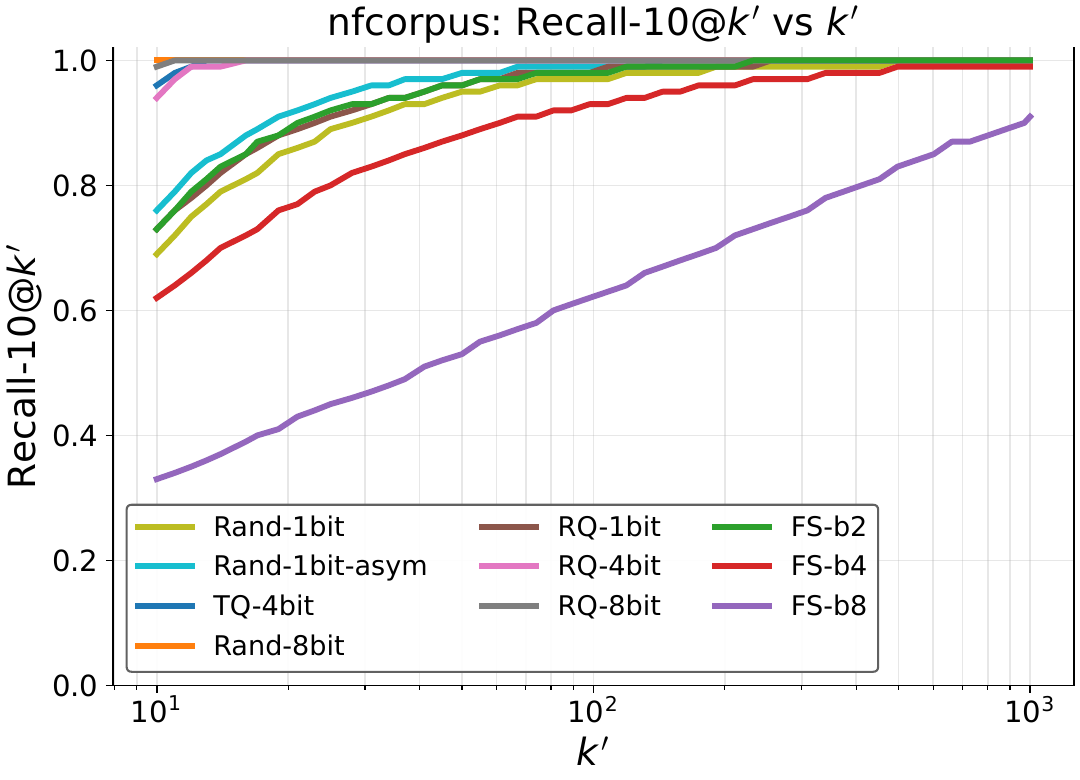}{0.235\linewidth}{nfcorpus}
    \hfill
    \appendixplot{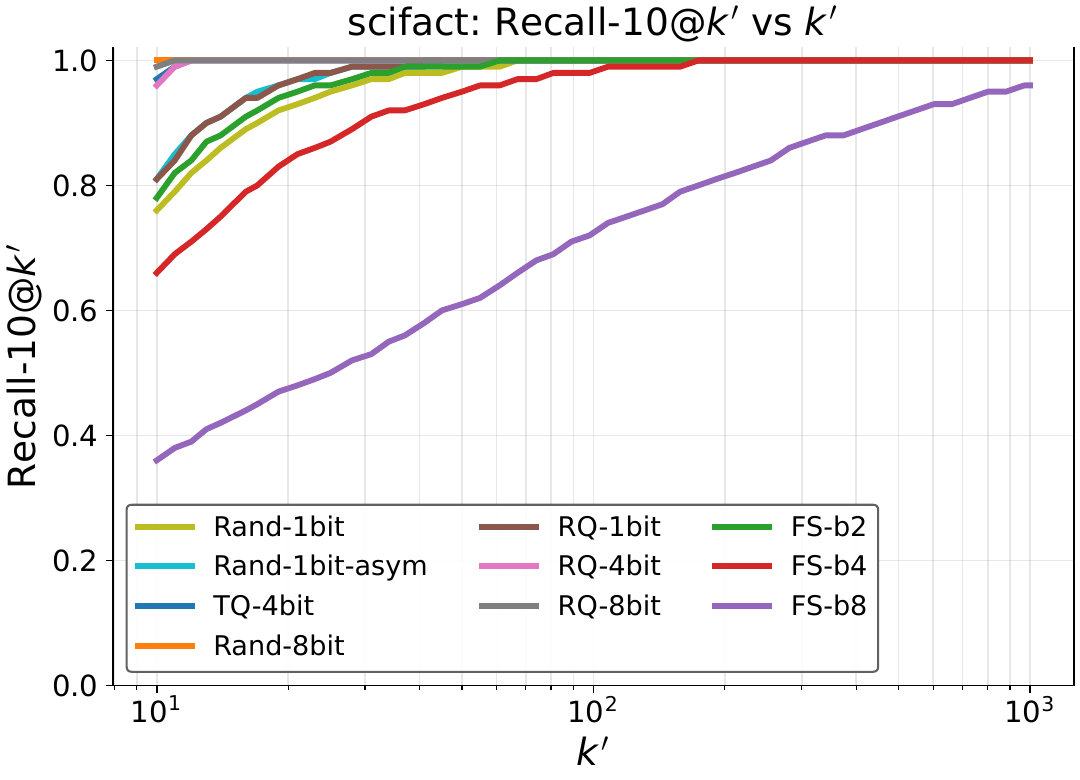}{0.235\linewidth}{scifact}
    \hfill
    \appendixplot{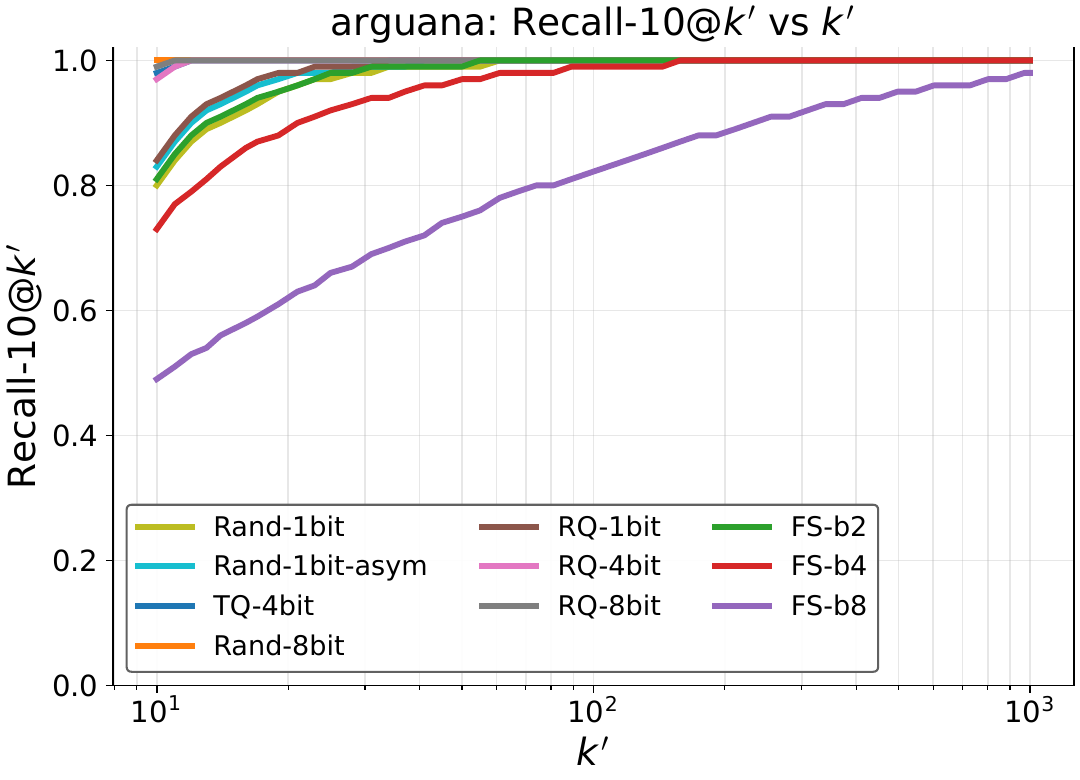}{0.235\linewidth}{arguana}
    \hfill
    \appendixplot{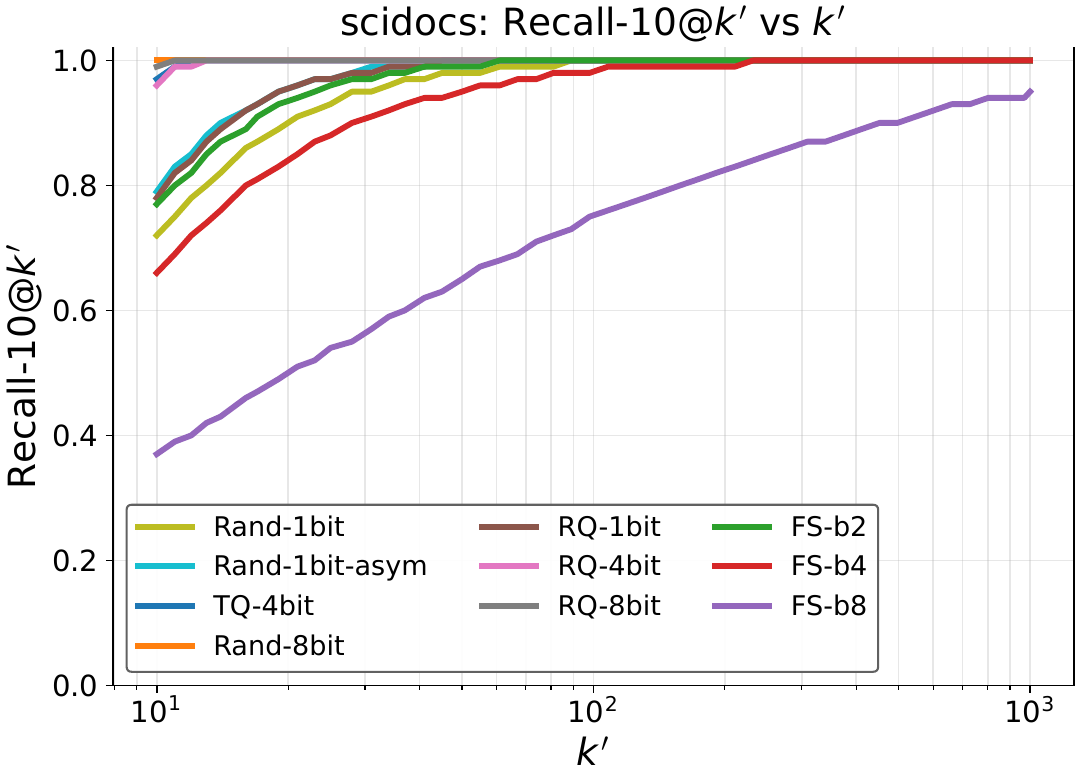}{0.235\linewidth}{scidocs}

    \vspace{0.75em}
    \appendixplot{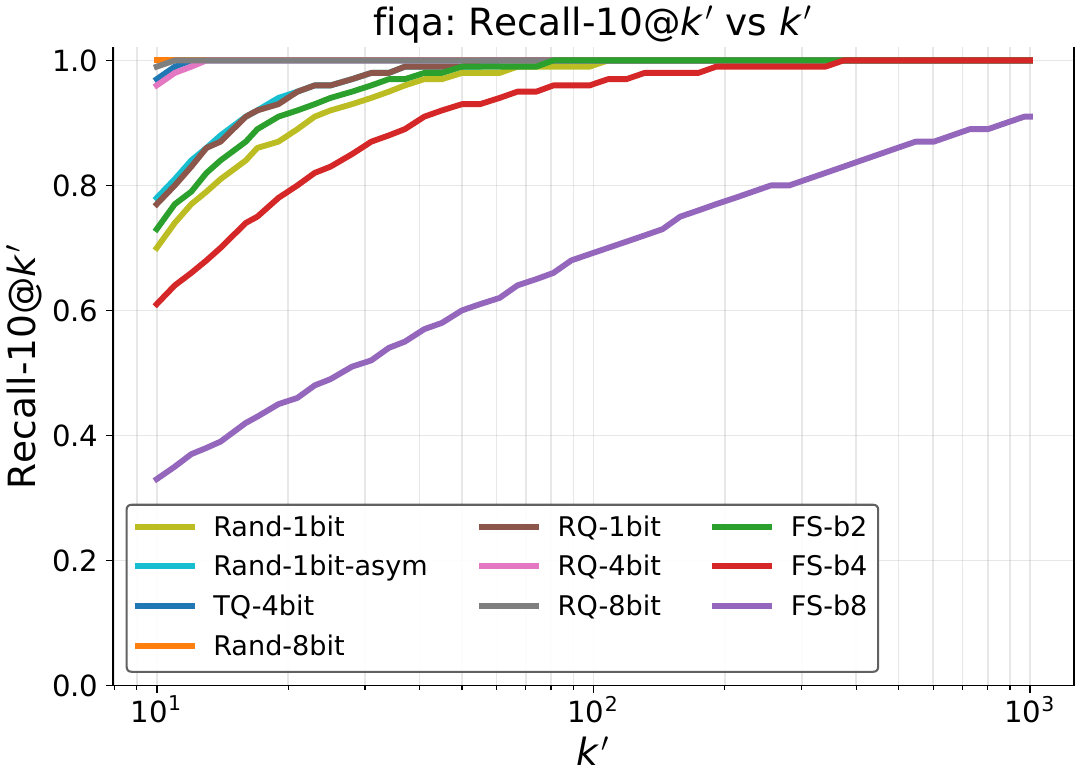}{0.30\linewidth}{fiqa}
    \hfill
    \appendixplot{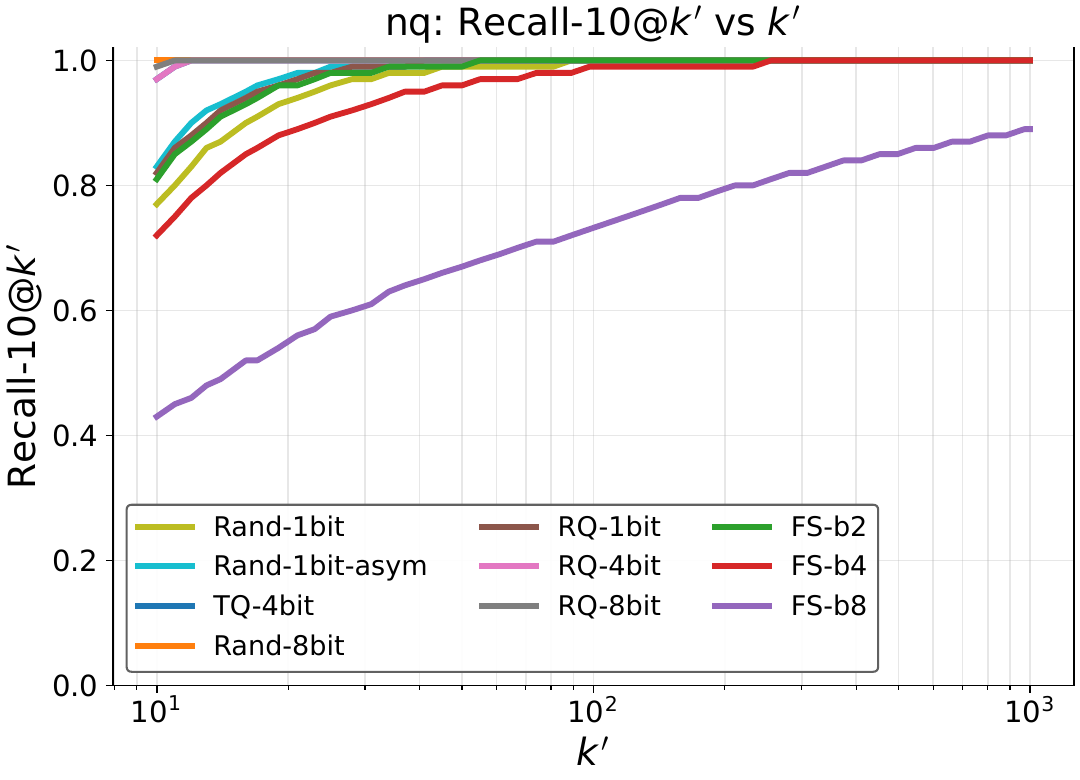}{0.30\linewidth}{nq}
    \hfill
    \appendixplot{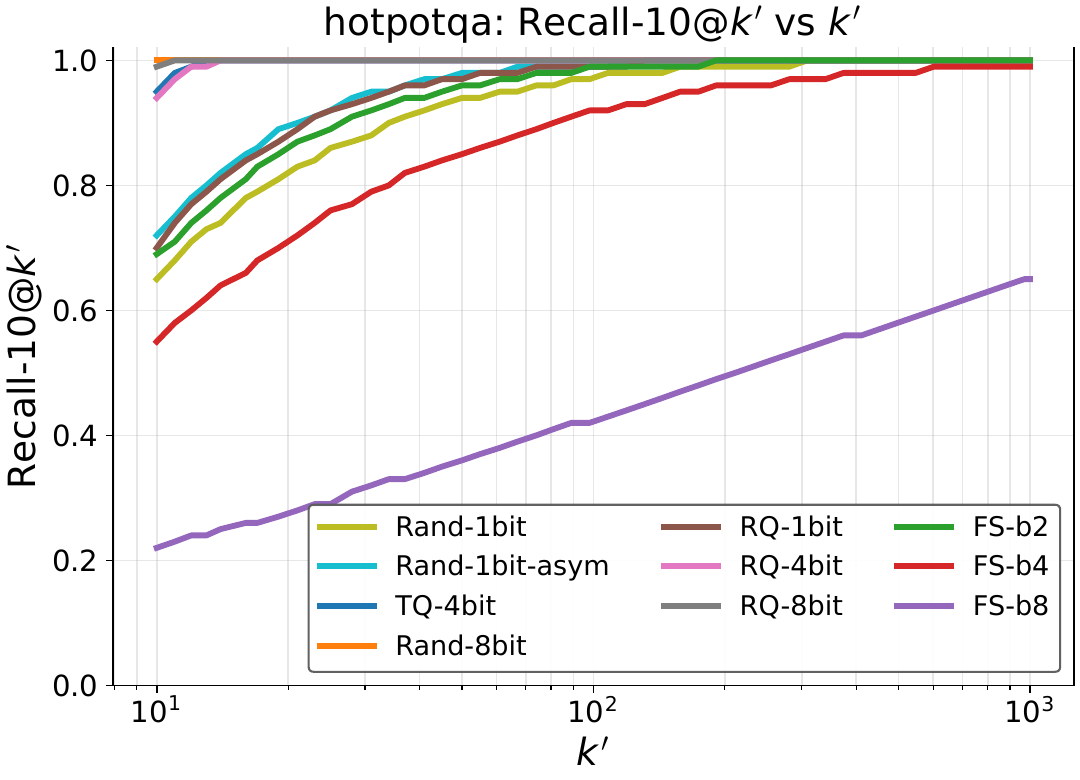}{0.30\linewidth}{hotpotqa}
    \caption{\small Multi-vector over-retrieval quality across datasets. Each panel shows recall as a function of the candidate budget $k'$.}
    \label{fig:app-multi-vector-overretrieve}
\end{figure*}

\myparagraph{Multi-vector over-retrieval benchmark}
We next evaluate the same quantizers inside the multi-vector retrieval pipeline. Figure~\ref{fig:app-multi-vector-overretrieve} reports the resulting recall--over-retrieval curves.

An interesting observation is that the quantizers perform comparatively better
in the multi-vector setting than in the single-vector setting. One reason is
that $\maxsim$ aggregates many query-token contributions into a single
query-document score. As a result, errors in individual inner-product estimates
can partially smooth out at the level of the full point-cloud score. This makes
even relatively low-precision quantizers useful for coarse multi-vector scoring,
especially when the goal is to recover a sufficiently large candidate set rather
than exactly match the full-precision ordering.

\subsection{Query Compression}\label{app:query-compression}
Multi-vector scoring is linear in the number of query embeddings, and many
queries contain redundant token vectors. We therefore compress each query point
cloud before scoring. Given a query $Q$, both compression methods considered
below produce a partition $\mathcal{B}=\{B_1,\ldots,B_r\}$ of the query
embeddings, and replace each block $B_i$ by a single representative $\bar q_i := \sum_{q\in B_i} q.$
The compressed query is then $\bar Q=\{\bar q_1,\ldots,\bar q_r\}$. Using sums
rather than averages preserves the scale of the original MaxSim score: a block
with more query embeddings contributes proportionally more to the final score.
Larger compression thresholds produce fewer representatives and faster scoring,
but can reduce retrieval quality.

\myparagraph{Ball Carving}
Ball carving is the greedy query-compression method used by
MUVERA~\cite{dhulipala2024muvera}. Given a distance threshold $\tau$, the
algorithm repeatedly selects an unassigned query embedding $q$, forms a block
consisting of all currently unassigned embeddings within distance $\tau$ of
$q$, replaces this block by the sum of its vectors, and removes the block from
the active set. This process continues until all query embeddings have been
assigned. Thus, the number of compressed query vectors is not fixed in advance;
it is determined by the geometry of the query and the threshold $\tau$.

\myparagraph{Ward's Method}
We also consider compression based on Ward's agglomerative clustering~\cite{ward1963hierarchical}. Each
query embedding is initialized as a singleton cluster. We then repeatedly merge
the pair of clusters that causes the smallest increase in within-cluster
variance, stopping once the next merge exceeds a threshold $\tau$. As with ball
carving, each final cluster is represented by the sum of its embeddings. Ward
compression tends to form more balanced and coherent groups of query vectors,
and in our experiments gives a slightly better quality--compression tradeoff.

\begin{figure*}[t]
    \centering
    Ball Carving\\[0.25em]
    \appendixplot{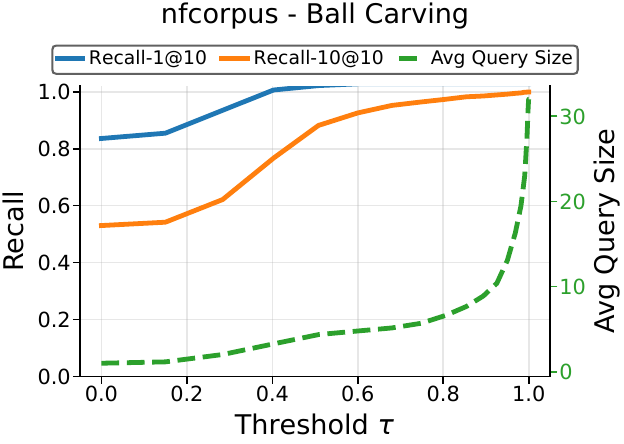}{0.19\linewidth}{nfcorpus}
    \hfill
    \appendixplot{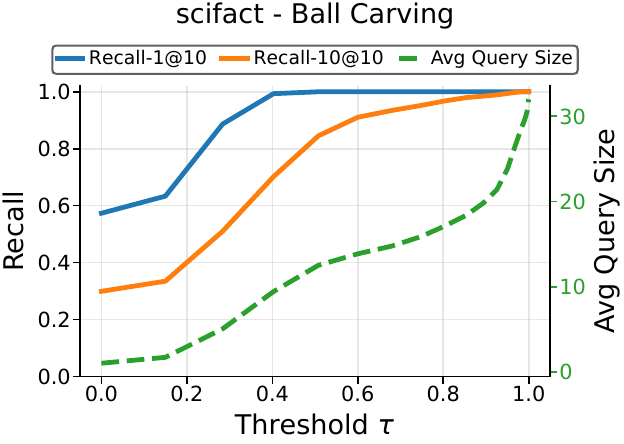}{0.19\linewidth}{scifact}
    \hfill
    \appendixplot{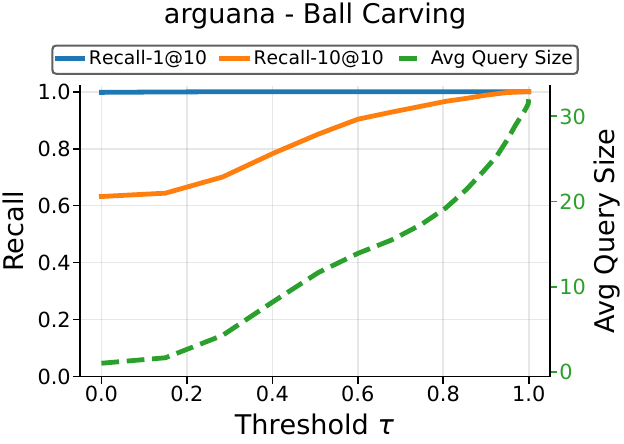}{0.19\linewidth}{arguana}
    \hfill
    \appendixplot{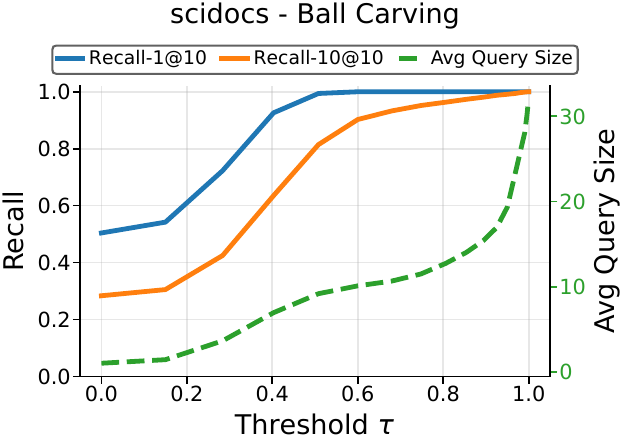}{0.19\linewidth}{scidocs}
    \hfill
    \appendixplot{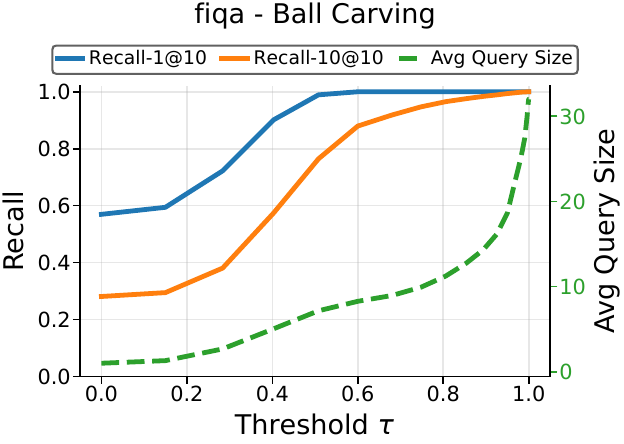}{0.19\linewidth}{fiqa}

    \vspace{0.75em}
    Ward's Method\\[0.25em]
    \appendixplot{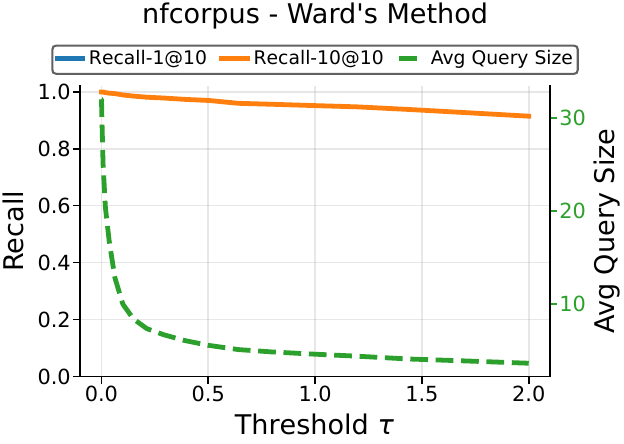}{0.19\linewidth}{nfcorpus}
    \hfill
    \appendixplot{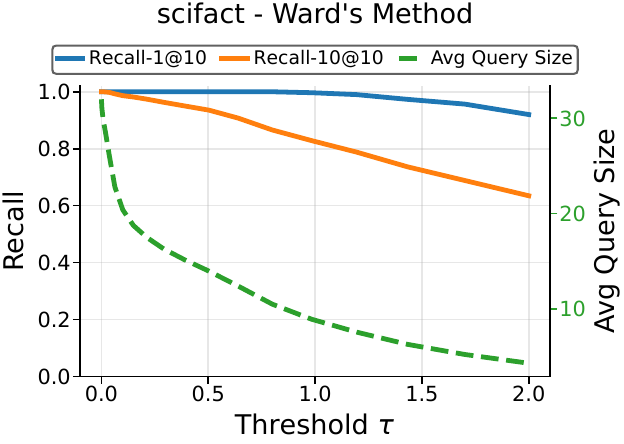}{0.19\linewidth}{scifact}
    \hfill
    \appendixplot{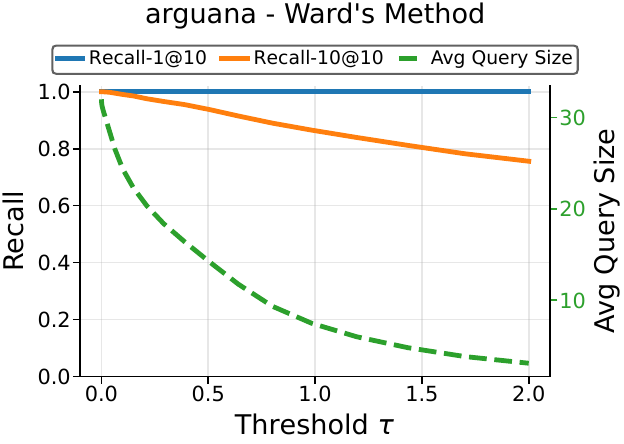}{0.19\linewidth}{arguana}
    \hfill
    \appendixplot{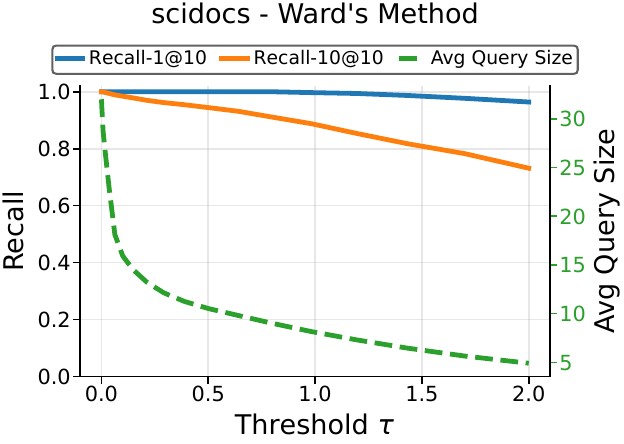}{0.19\linewidth}{scidocs}
    \hfill
    \appendixplot{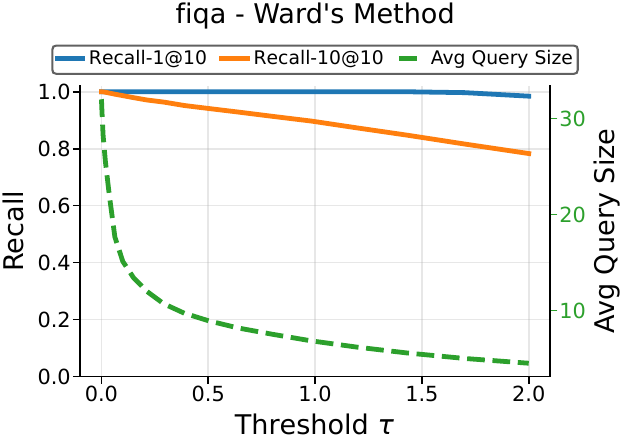}{0.19\linewidth}{fiqa}
    \caption{\small Query-compression quality for Ball Carving (top row) and Ward's method (bottom row). Each panel plots retrieval quality as the compression threshold varies.}
    \label{fig:query-compression}
\end{figure*}

\myparagraph{Discussion}
Figure~\ref{fig:query-compression} compares the two compression methods as the
threshold is varied. Both methods are effective: moderate compression reduces
the number of query embeddings substantially with little effect on recall. We
use Ward's method with threshold $0.5$ by default, since it is slightly faster in our
implementation and gives comparable or better recall across the operating
points we consider. The cost of query compression itself is negligible compared to the overall search time; its main effect is to reduce the cost of the subsequent multi-vector scoring kernels. Furthermore, most of the loss in quality here can be recovered with reranking, allowing us to compress more aggressively.

\subsection{Extended Latency Experiments}\label{app:latency}
We present the latency vs recall plots on all of the datasets here; see Figures~\ref{fig:app-latency-nfcorpus},\ref{fig:app-latency-scifact},\ref{fig:app-latency-arguana},\ref{fig:app-latency-scidocs},\ref{fig:app-latency-fiqa},\ref{fig:app-latency-nq},\ref{fig:app-latency-hotpotqa},\ref{fig:app-latency-msmarco}, and~\ref{fig:app-latency-lotte}. See Tables~\ref{tab:recall10-speedups-geomean} and~\ref{tab:recall100-speedups-geomean} for a summary of the speedups obtained.

\appendixevaluationplot{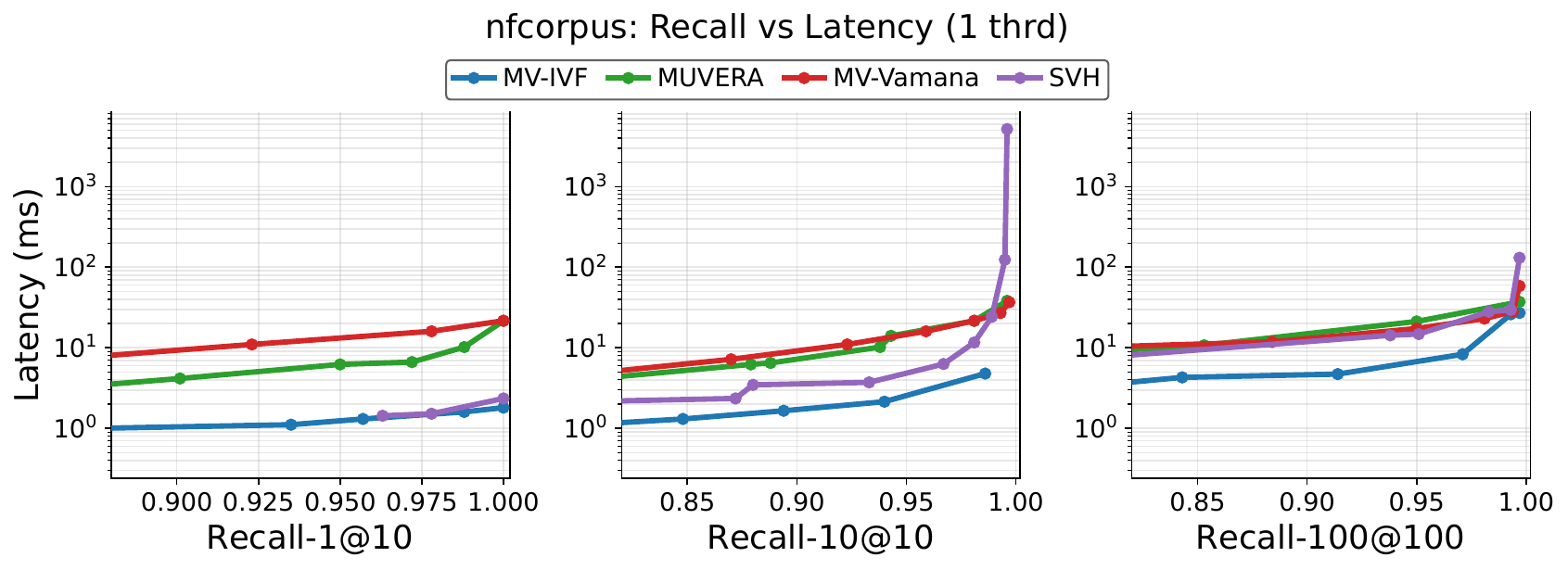}{Latency}{nfcorpus}{fig:app-latency-nfcorpus}
\appendixevaluationplot{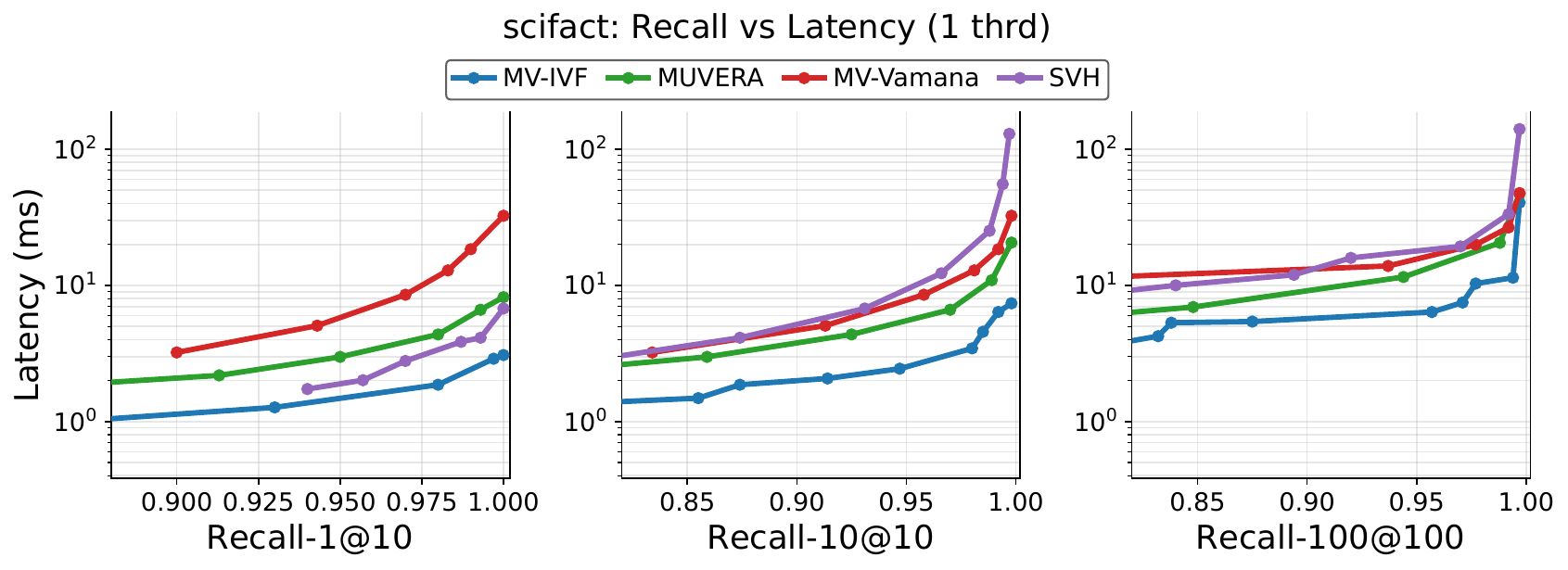}{Latency}{scifact}{fig:app-latency-scifact}
\appendixevaluationplot{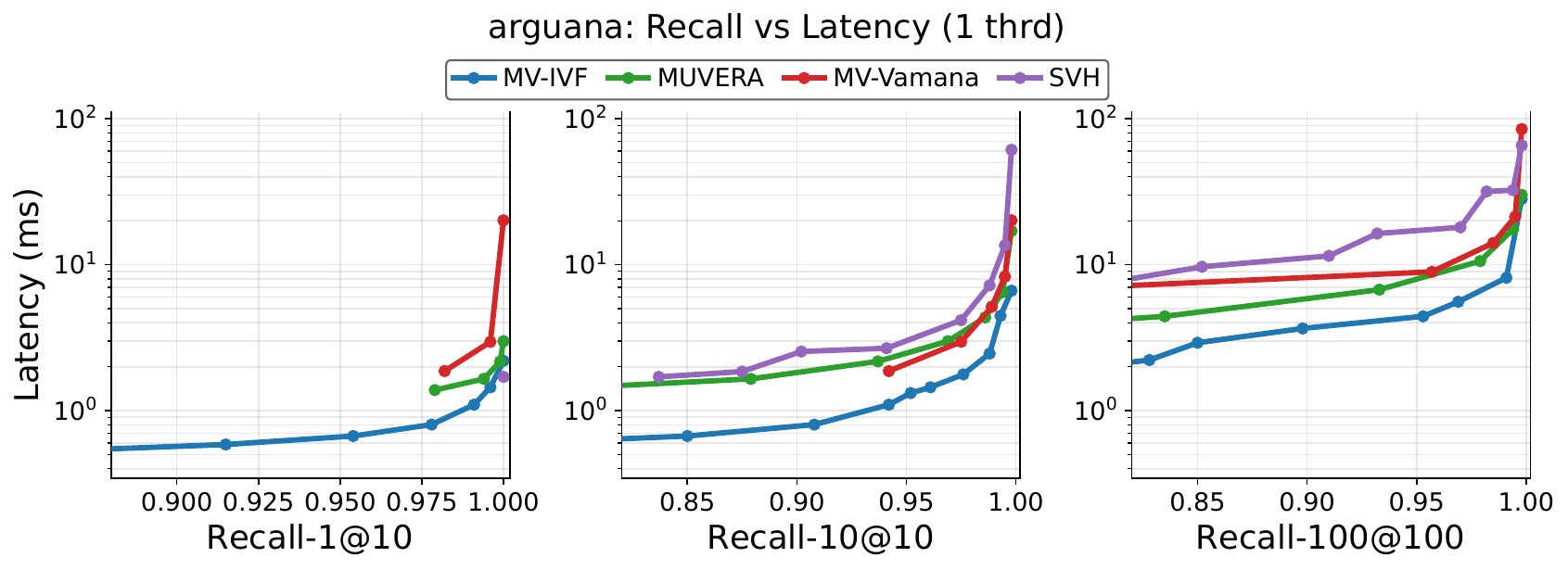}{Latency}{arguana}{fig:app-latency-arguana}
\appendixevaluationplot{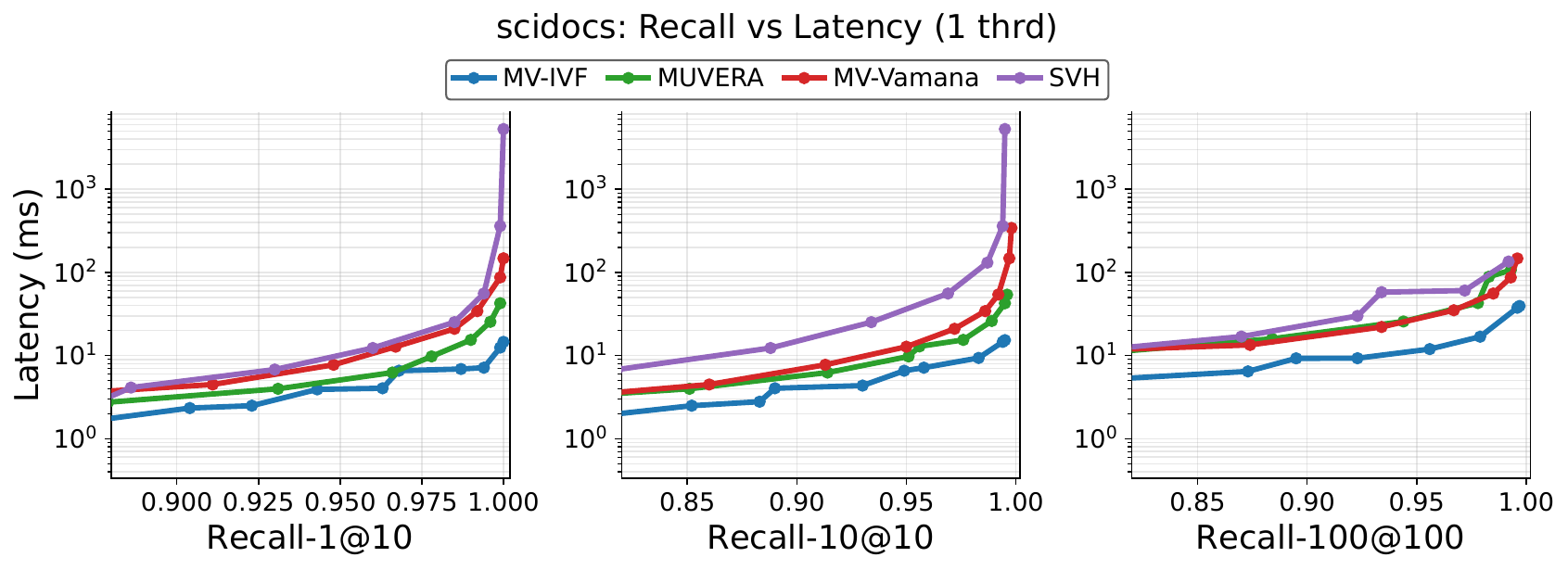}{Latency}{scidocs}{fig:app-latency-scidocs}
\appendixevaluationplot{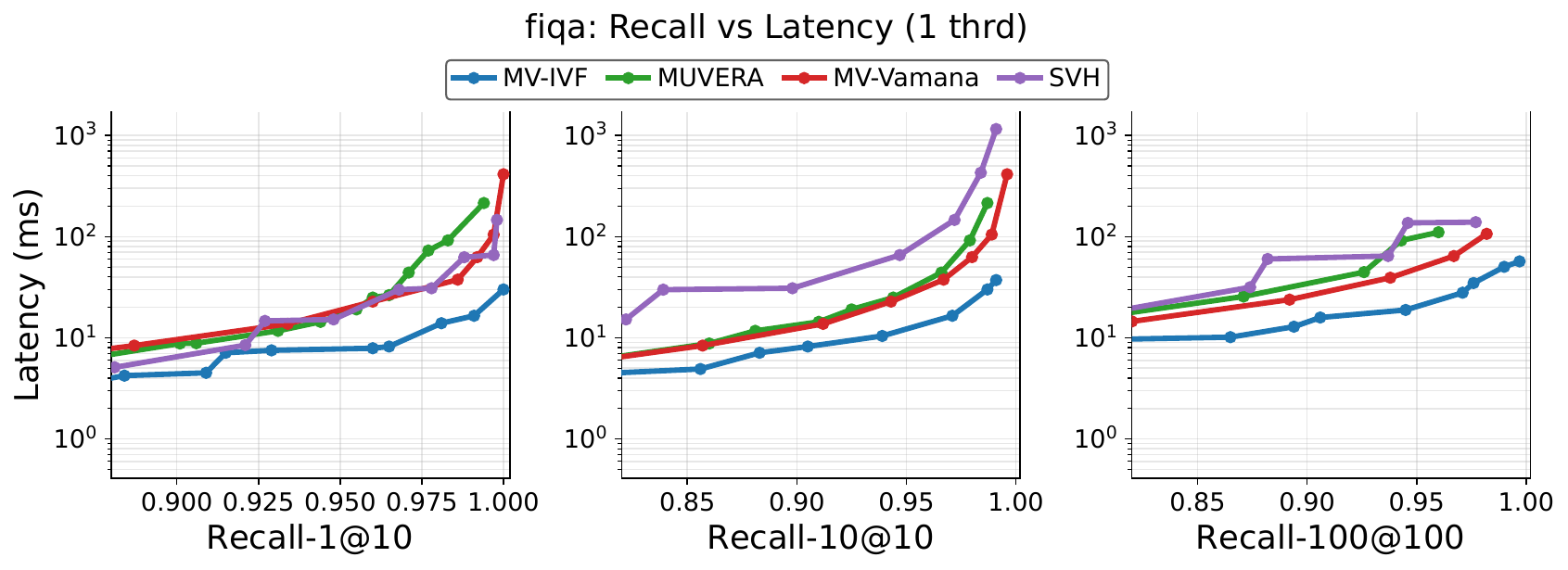}{Latency}{fiqa}{fig:app-latency-fiqa}
\appendixevaluationplot{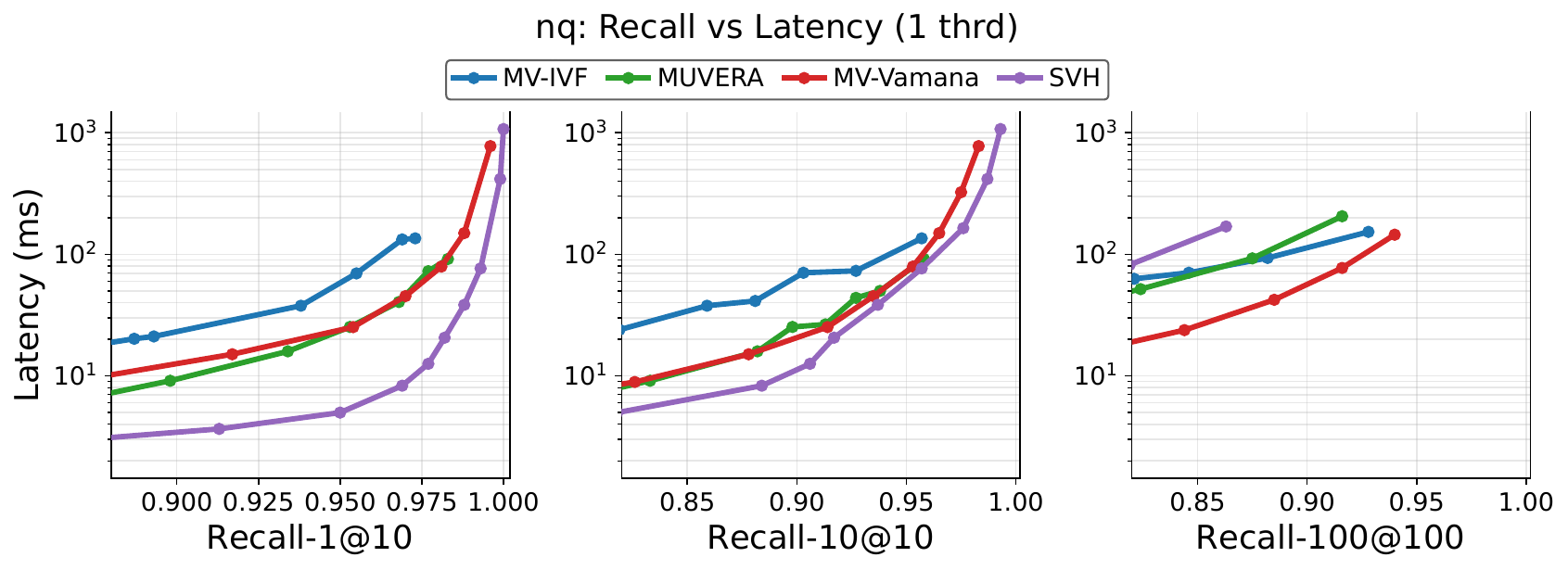}{Latency}{nq}{fig:app-latency-nq}
\appendixevaluationplot{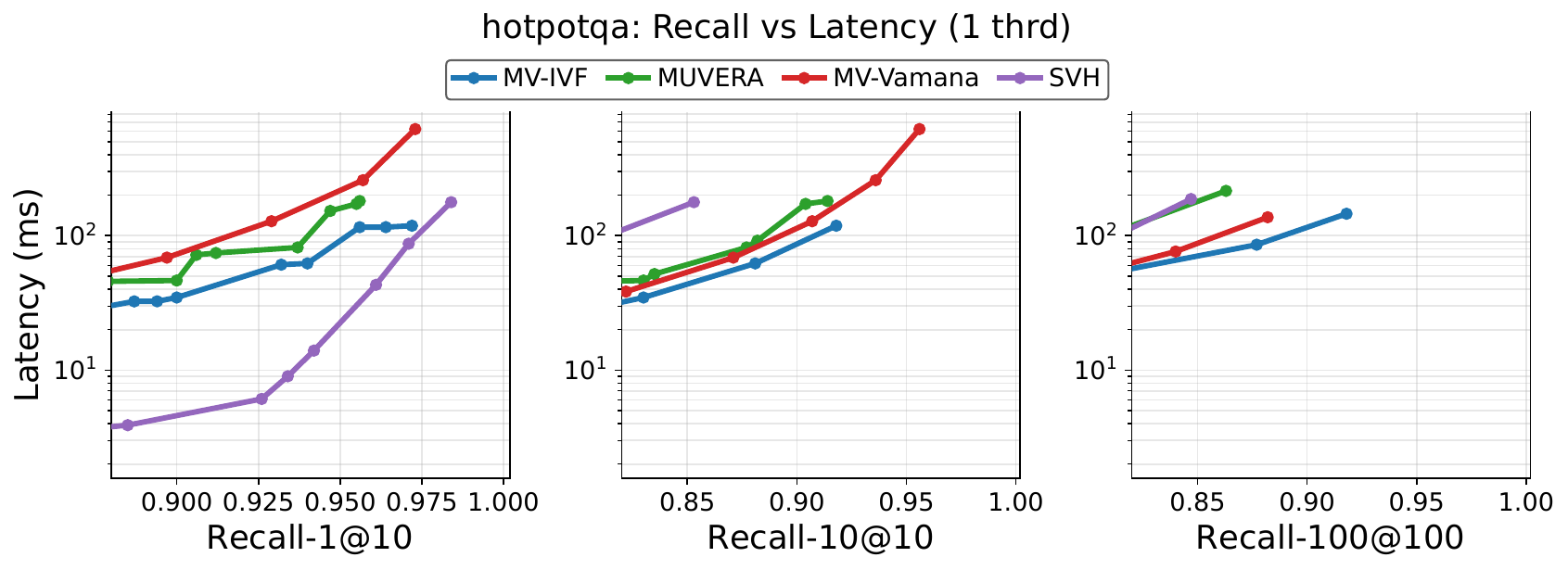}{Latency}{hotpotqa}{fig:app-latency-hotpotqa}
\appendixevaluationplot{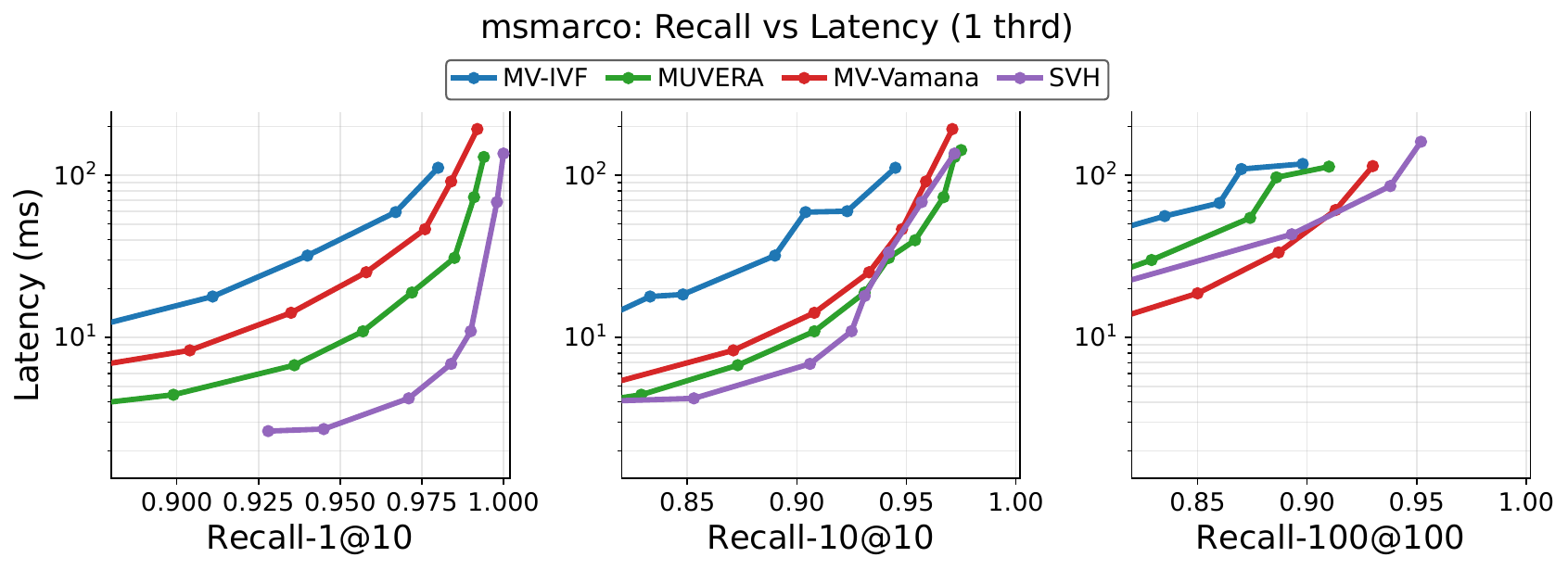}{Latency}{msmarco}{fig:app-latency-msmarco}
\appendixevaluationplot{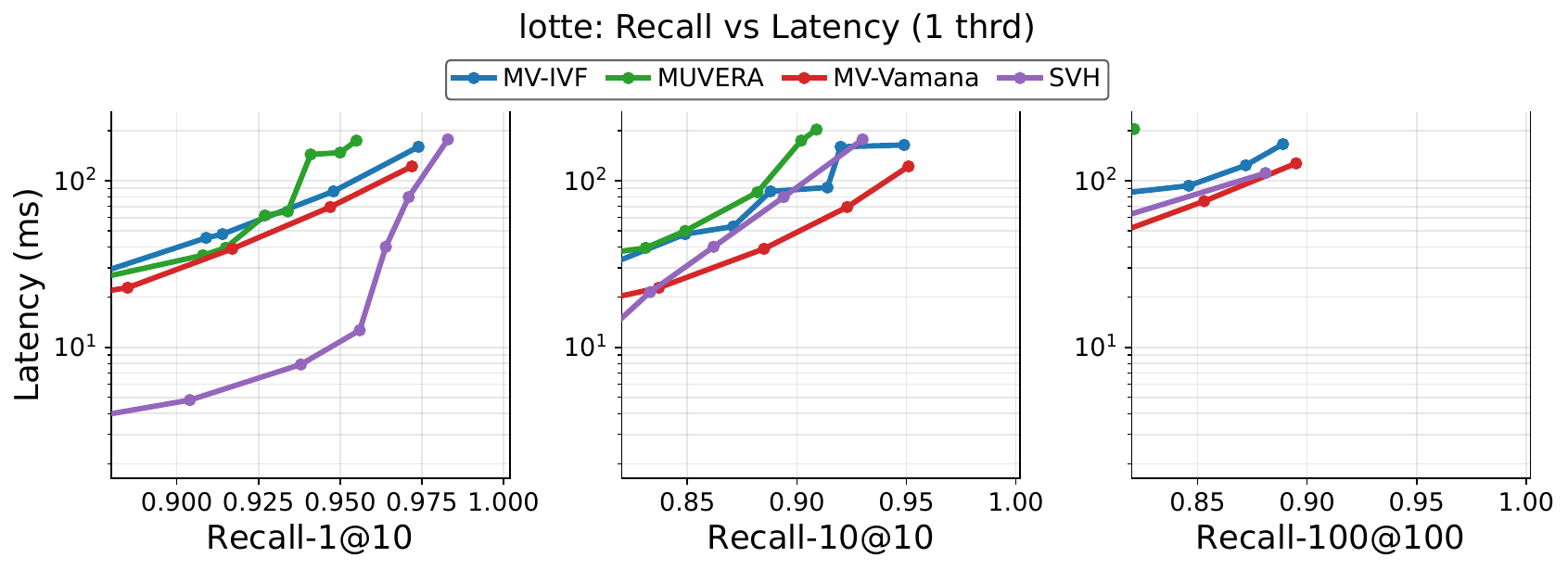}{Latency}{lotte}{fig:app-latency-lotte}

\subsection{Extended Throughput Experiments}\label{app:throughput}
We present the batch throughput vs recall plots on all of the datasets here; see Figures~\ref{fig:app-throughput-nfcorpus},\ref{fig:app-throughput-scifact},\ref{fig:app-throughput-arguana},\ref{fig:app-throughput-scidocs},\ref{fig:app-throughput-fiqa},\ref{fig:app-throughput-nq},\ref{fig:app-throughput-hotpotqa},\ref{fig:app-throughput-msmarco}, and~\ref{fig:app-throughput-lotte}. See Tables~\ref{tab:recall10-speedups-geomean} and~\ref{tab:recall100-speedups-geomean} for a summary of the speedups obtained.

\appendixevaluationplot{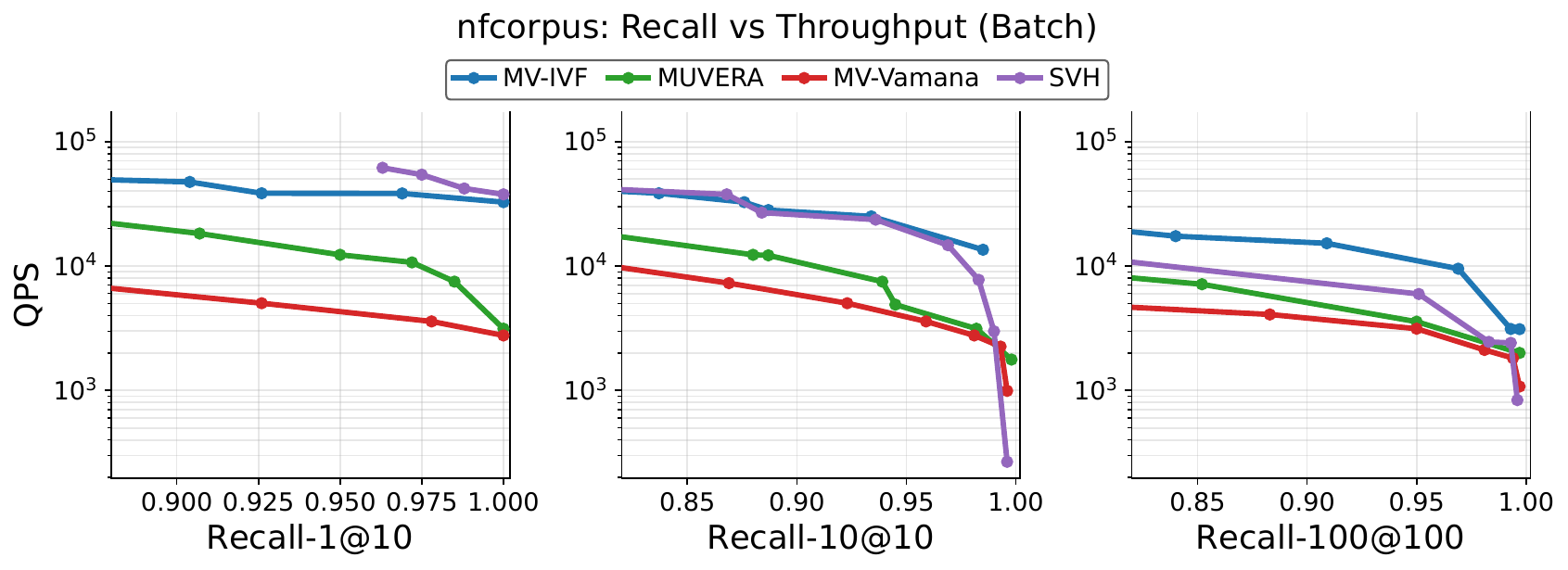}{Throughput}{nfcorpus}{fig:app-throughput-nfcorpus}
\appendixevaluationplot{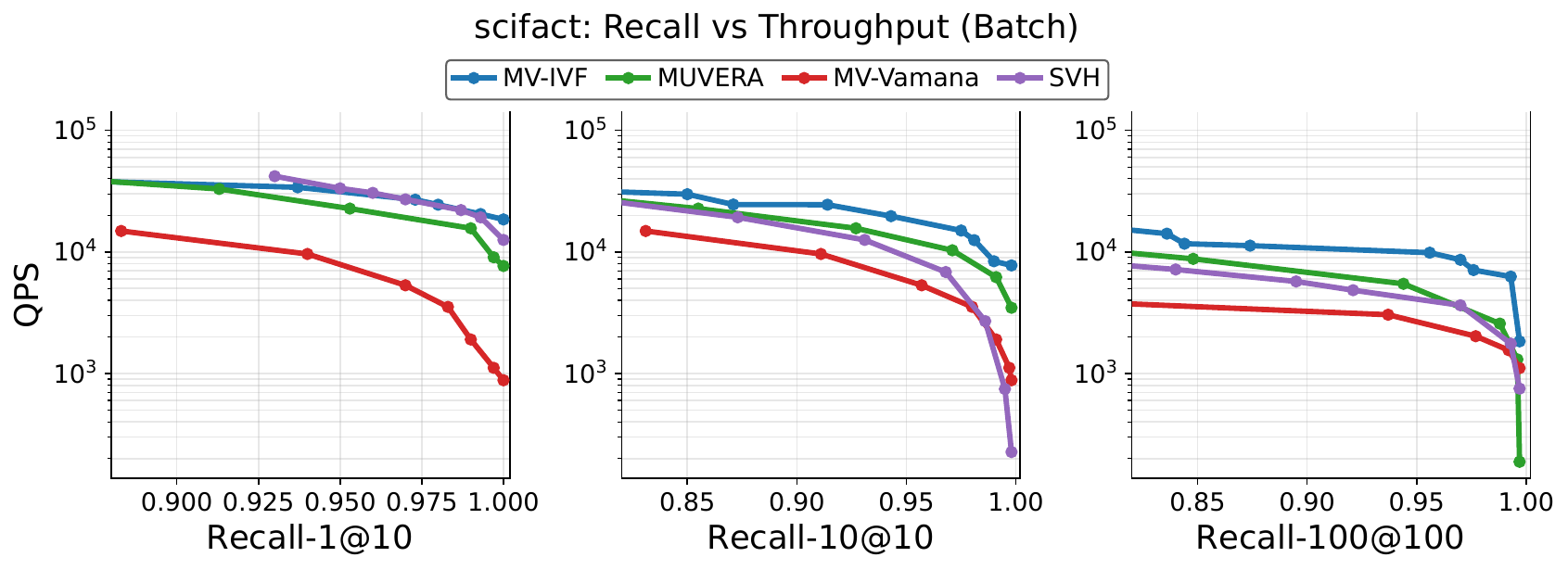}{Throughput}{scifact}{fig:app-throughput-scifact}
\appendixevaluationplot{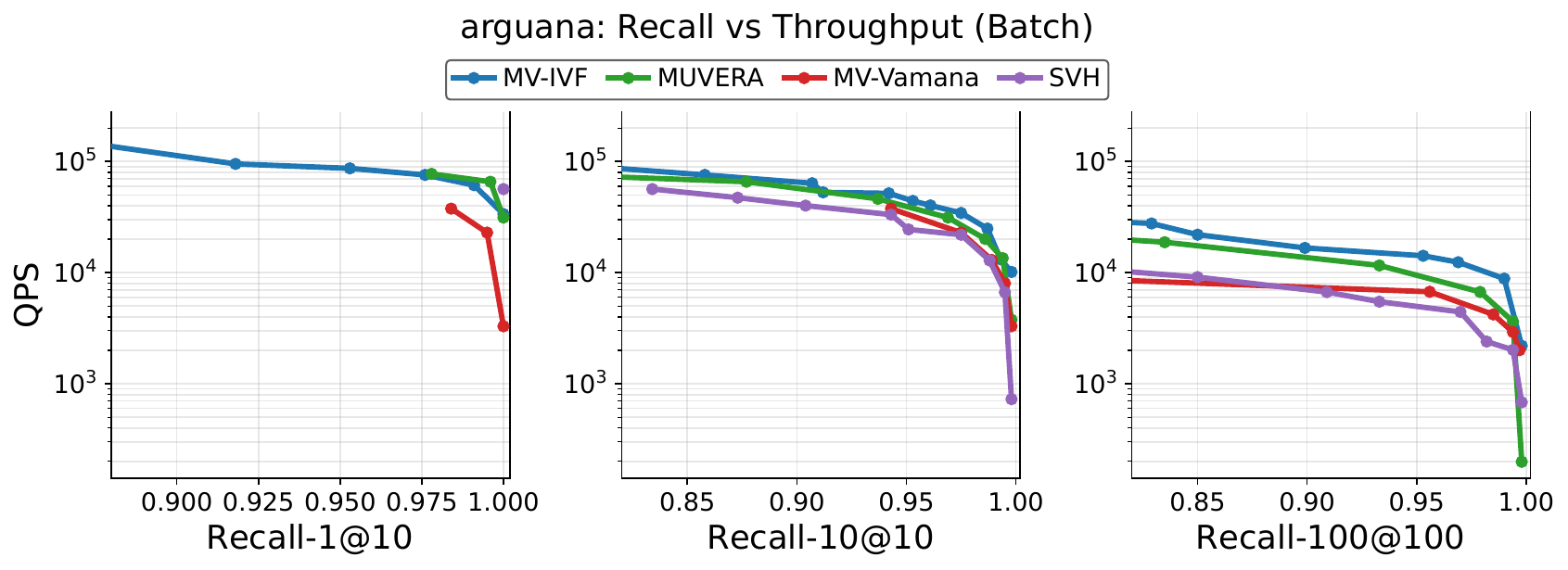}{Throughput}{arguana}{fig:app-throughput-arguana}
\appendixevaluationplot{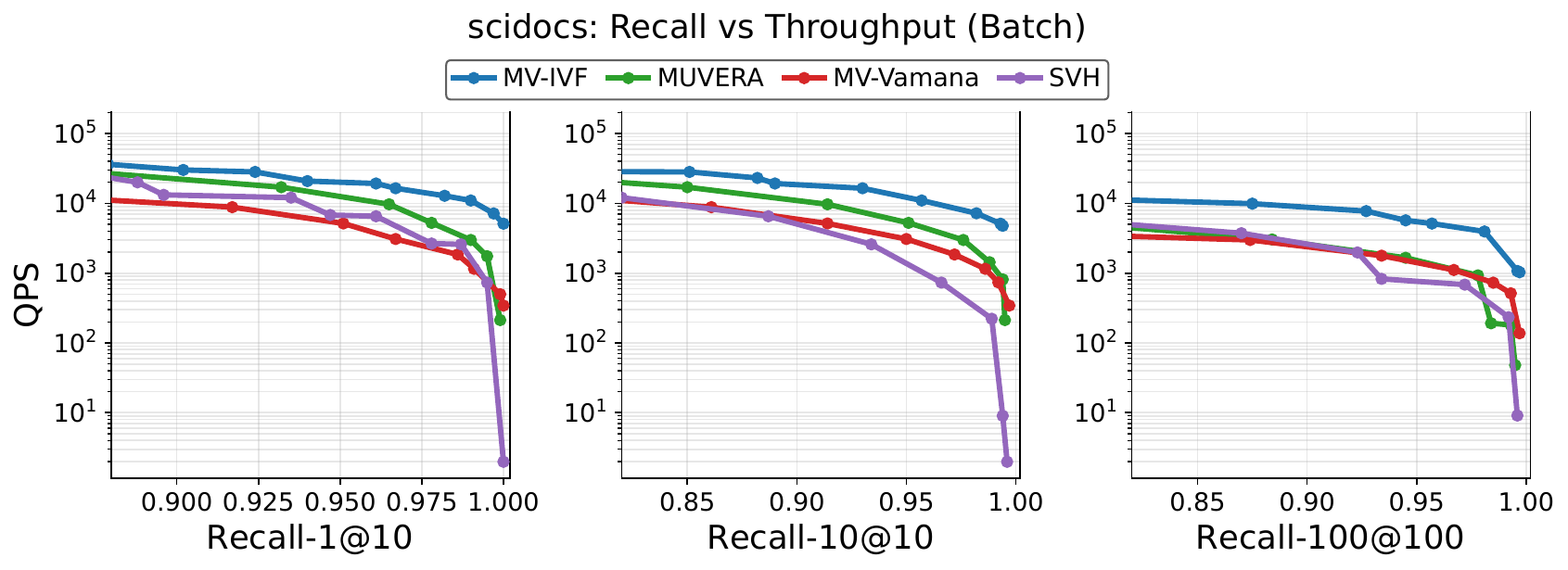}{Throughput}{scidocs}{fig:app-throughput-scidocs}
\appendixevaluationplot{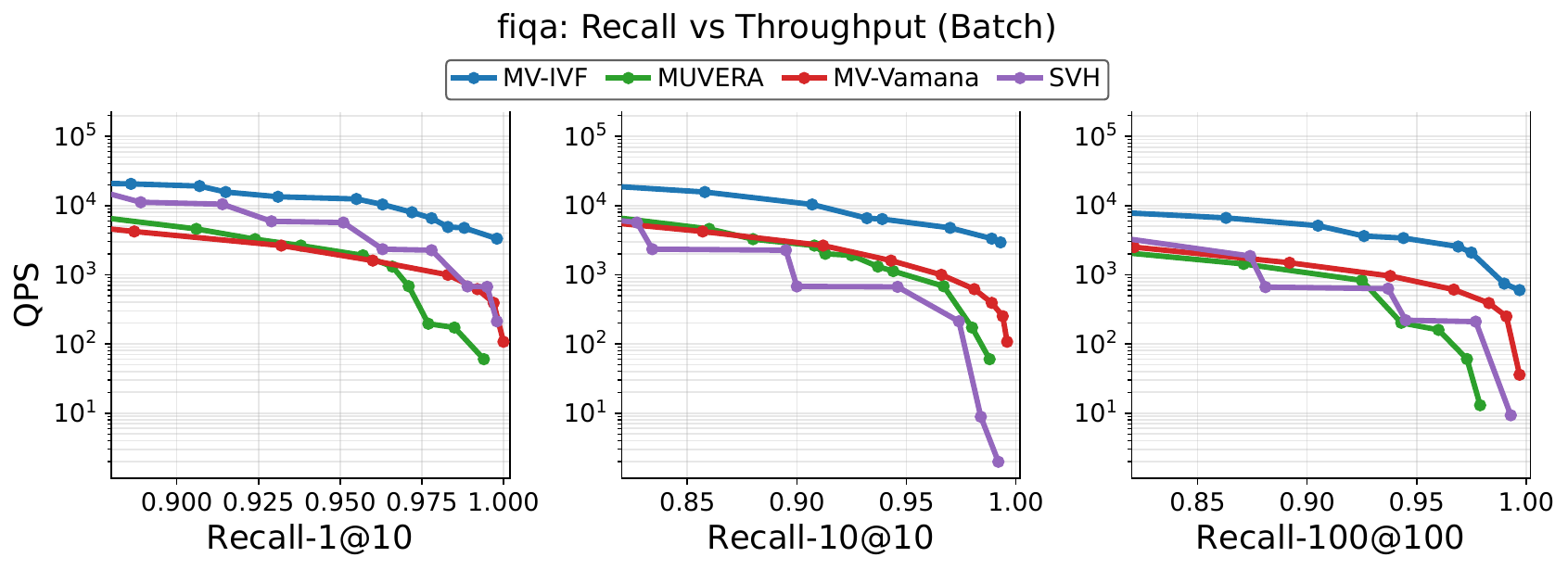}{Throughput}{fiqa}{fig:app-throughput-fiqa}
\appendixevaluationplot{throughput_quora.pdf}{Throughput}{quora}{fig:app-throughput-quora}
\appendixevaluationplot{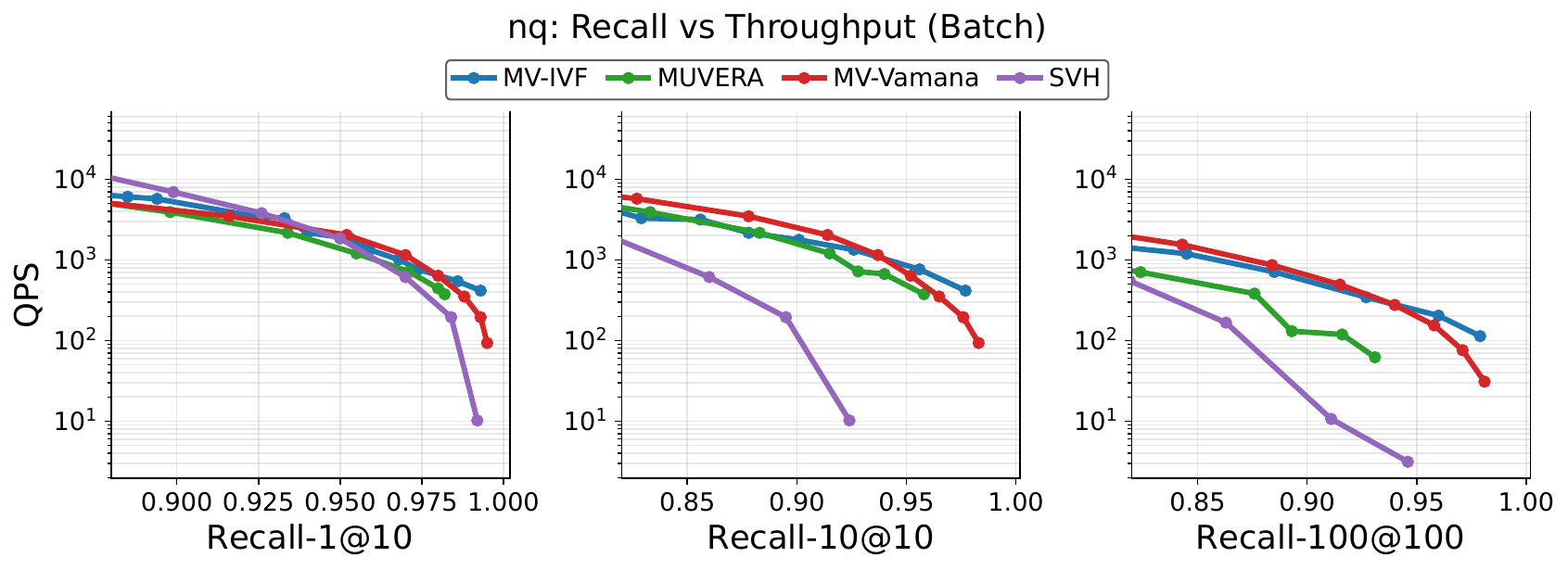}{Throughput}{nq}{fig:app-throughput-nq}
\appendixevaluationplot{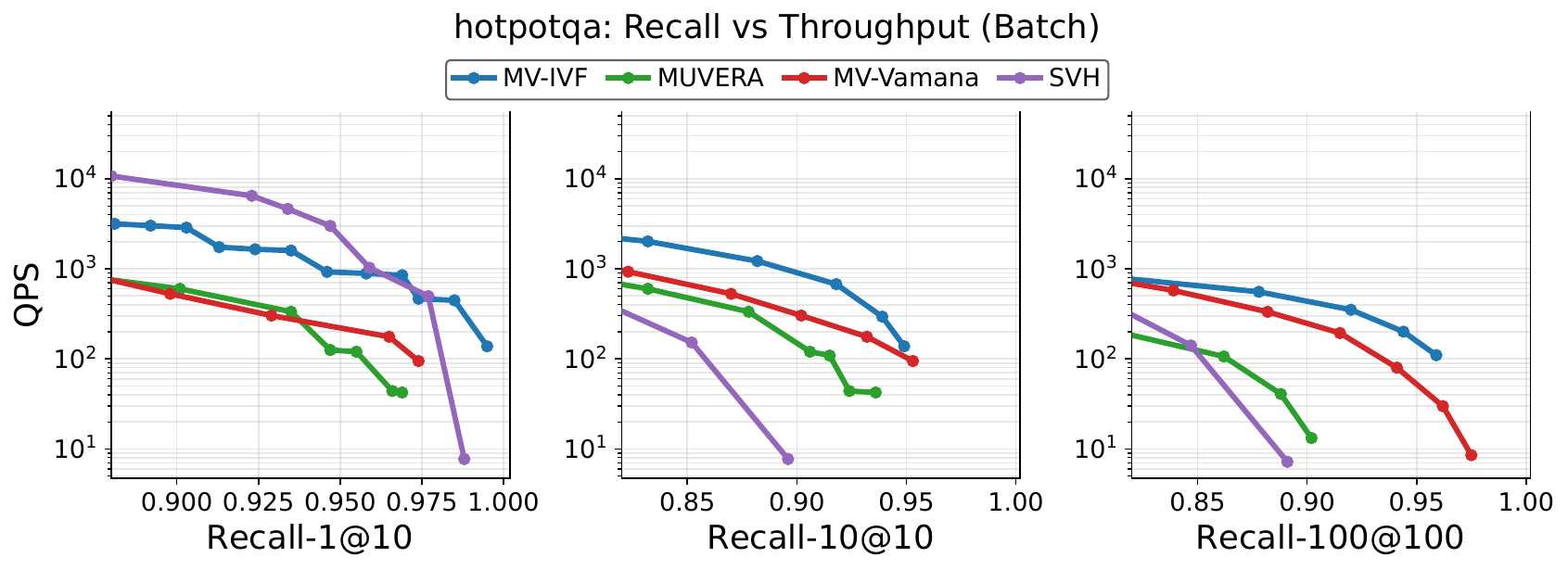}{Throughput}{hotpotqa}{fig:app-throughput-hotpotqa}
\appendixevaluationplot{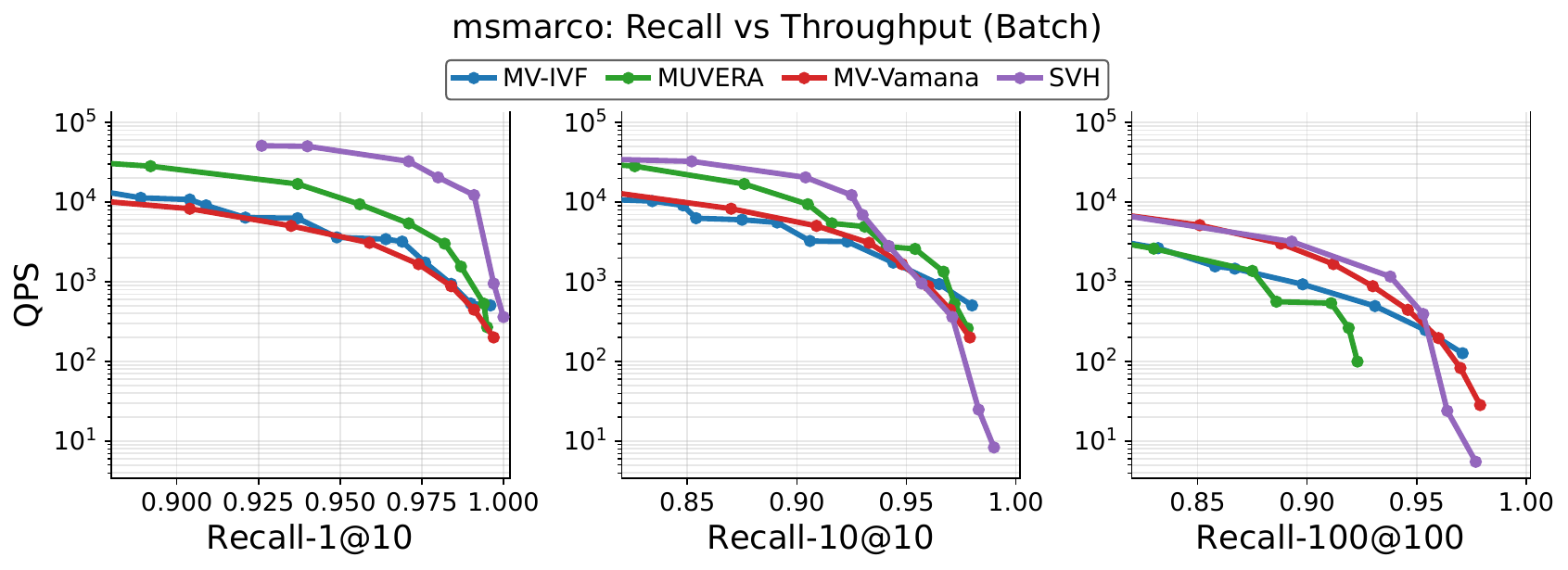}{Throughput}{msmarco}{fig:app-throughput-msmarco}
\appendixevaluationplot{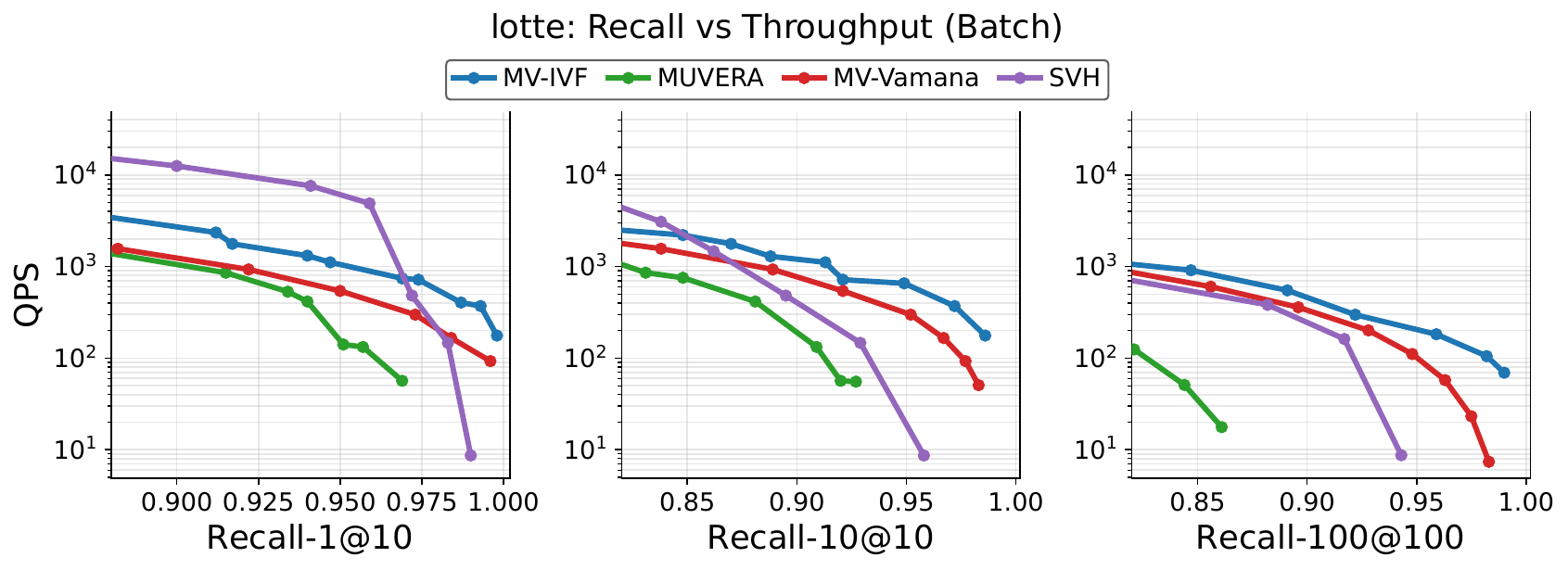}{Throughput}{lotte}{fig:app-throughput-lotte}

\begin{table*}[t]
\centering
\footnotesize
\setlength{\tabcolsep}{4pt}
\begin{tabular}{llccc}
\toprule
Regime & Baseline & 90\% Recall-$10$@$10$ & 95\% Recall-$10$@$10$ & 99\% Recall-$10$@$10$ \\
\midrule
\multirow{3}{*}{Latency}
  & MUVERA    & $1.33\times$ / $4.76\times$ (9) & $1.91\times$ / $4.51\times$ (6) & $2.09\times$ / $2.92\times$ (3) \\
  & SVH       & $1.55\times$ / $8.03\times$ (8) & $2.89\times$ / $8.86\times$ (6) & $12.46\times$ / $31.09\times$ (4) \\
  & MV-Vamana & $1.22\times$ / $5.13\times$ (9) & $1.88\times$ / $3.35\times$ (6) & $3.30\times$ / $4.98\times$ (4) \\
\midrule
\multirow{3}{*}{Throughput}
  & MUVERA    & $2.19\times$ / $8.40\times$ (9) & $1.92\times$ / $6.97\times$ (7) & $2.10\times$ / $6.28\times$ (3) \\
  & SVH       & $4.01\times$ / $173.53\times$ (8) & $4.71\times$ / $42.85\times$ (7) & $67.49\times$ / $1482.60\times$ (4) \\
  & MV-Vamana & $2.07\times$ / $4.99\times$ (9) & $2.20\times$ / $4.77\times$ (8) & $5.05\times$ / $11.65\times$ (4) \\
\bottomrule
\end{tabular}
\caption{\small \textbf{MVIVF speedups.} Each cell reports geometric-mean speedup / maximum speedup over the baseline, with the number of datasets compared in parentheses. The latency/throughput is computed by taking the best value with Recall-$10$@$10$ at least the fixed value.}
\label{tab:recall10-speedups-geomean}
\end{table*}

\begin{table*}[t]
\centering
\footnotesize
\setlength{\tabcolsep}{4pt}
\begin{tabular}{llccc}
\toprule
Regime & Baseline & 90\% Recall-$100$@$100$ & 95\% Recall-$100$@$100$ & 99\% Recall-$100$@$100$ \\
\midrule
\multirow{3}{*}{Latency}
  & MUVERA    & $2.25\times$ / $4.50\times$ (6) & $3.08\times$ / $3.96\times$ (5) & $2.45\times$ / $4.13\times$ (4) \\
  & SVH       & $3.03\times$ / $4.06\times$ (5) & $3.54\times$ / $5.06\times$ (5) & $2.63\times$ / $3.99\times$ (4) \\
  & MV-Vamana & $1.91\times$ / $3.68\times$ (6) & $2.45\times$ / $3.12\times$ (5) & $1.98\times$ / $2.62\times$ (4) \\
\midrule
\multirow{3}{*}{Throughput}
  & MUVERA    & $3.43\times$ / $26.55\times$ (8) & $4.53\times$ / $16.00\times$ (5) & $3.22\times$ / $5.93\times$ (4) \\
  & SVH       & $3.12\times$ / $32.24\times$ (8) & $3.04\times$ / $12.16\times$ (6) & $5.95\times$ / $80.08\times$ (5) \\
  & MV-Vamana & $1.96\times$ / $5.34\times$ (9) & $2.83\times$ / $4.86\times$ (9) & $2.65\times$ / $4.03\times$ (5) \\
\bottomrule
\end{tabular}
\caption{\small \textbf{MVIVF speedups.} Each cell reports geometric-mean speedup / maximum speedup over the baseline, with the number of datasets compared in parentheses. The latency/throughput is computed by taking the best value with Recall-$100$@$100$ at least the fixed value.}
\label{tab:recall100-speedups-geomean}
\end{table*}


\subsection{Multi-thread Latency Experiments}\label{app:multi_latency}
We also present the multi-threaded latency vs recall plots, i.e. latency when single-queries run on all cores. This measures the amount of intra-query parallelization in the algorithm.  See Figures~\ref{fig:app-multi-latency-nfcorpus},\ref{fig:app-multi-latency-scifact},\ref{fig:app-multi-latency-arguana},\ref{fig:app-multi-latency-scidocs}, and~\ref{fig:app-multi-latency-fiqa}.

\appendixevaluationplot{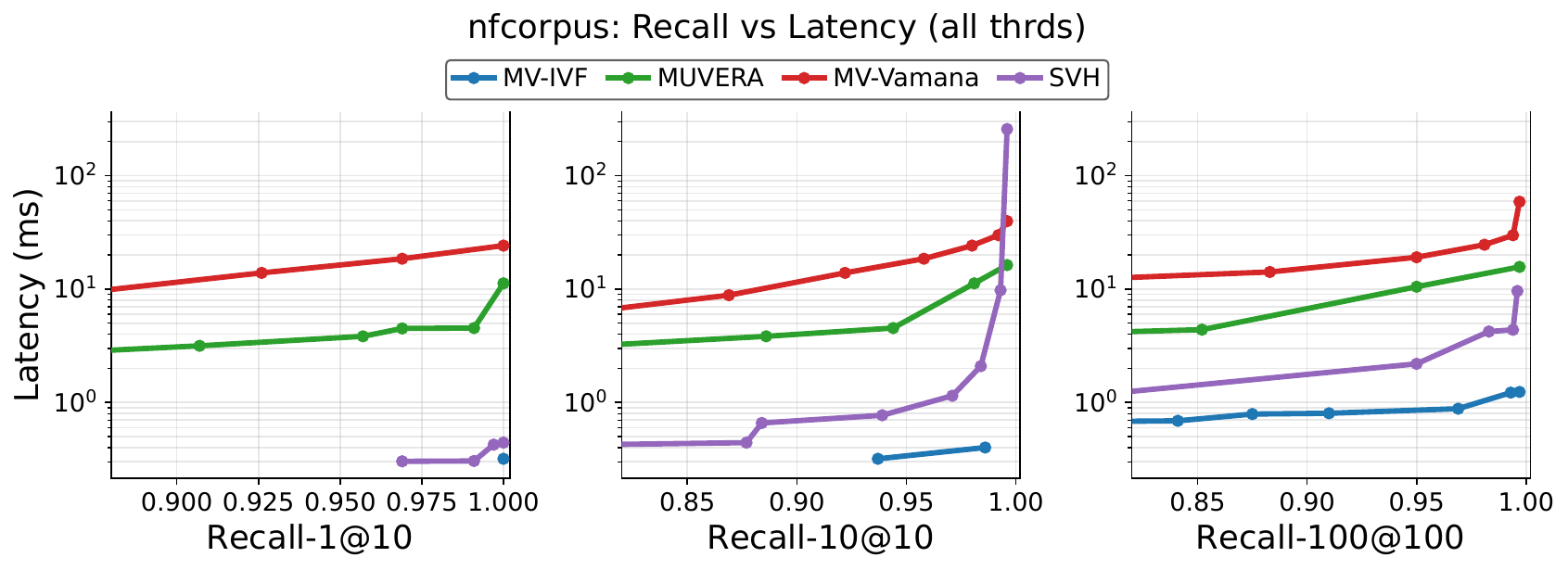}{Multi-thread latency}{nfcorpus}{fig:app-multi-latency-nfcorpus}
\appendixevaluationplot{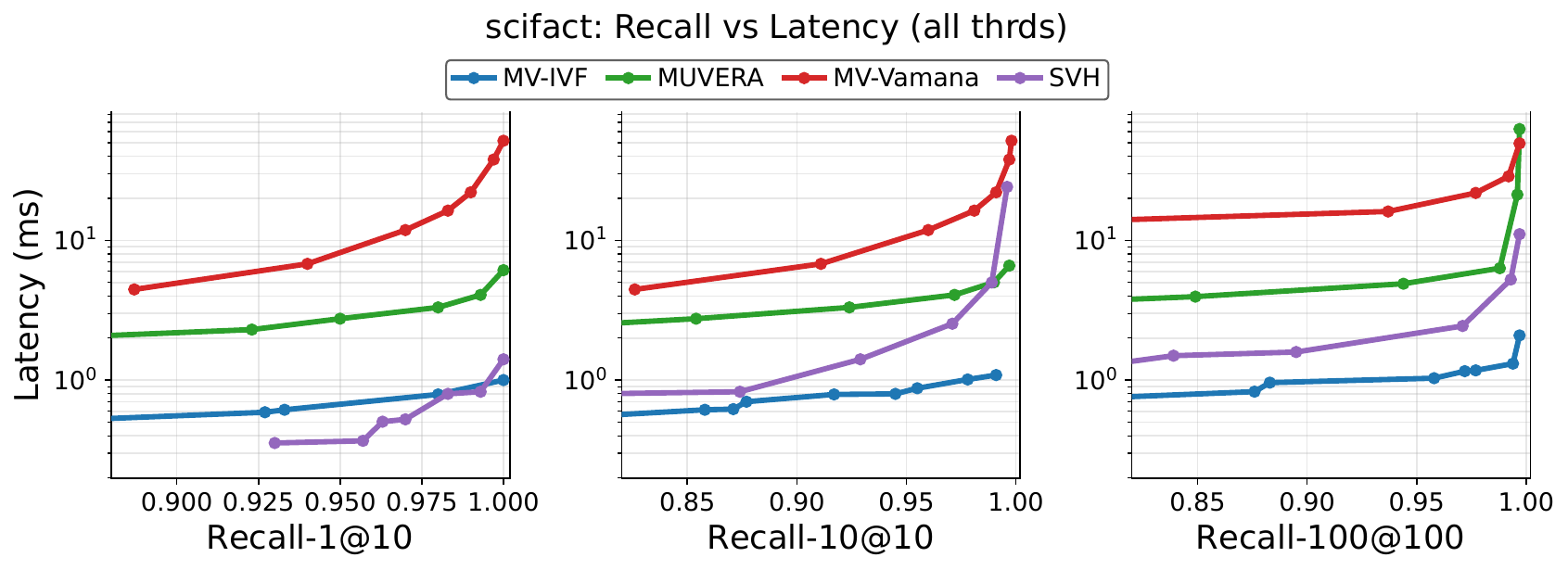}{Multi-thread latency}{scifact}{fig:app-multi-latency-scifact}
\appendixevaluationplot{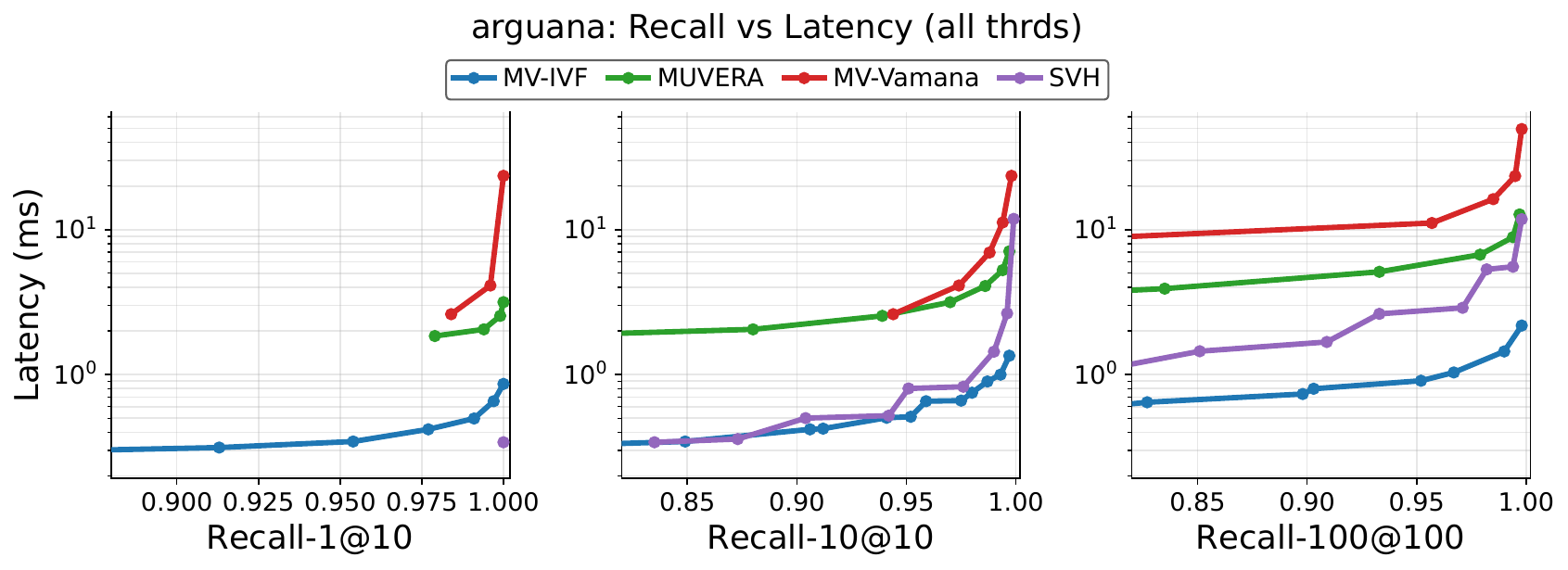}{Multi-thread latency}{arguana}{fig:app-multi-latency-arguana}
\appendixevaluationplot{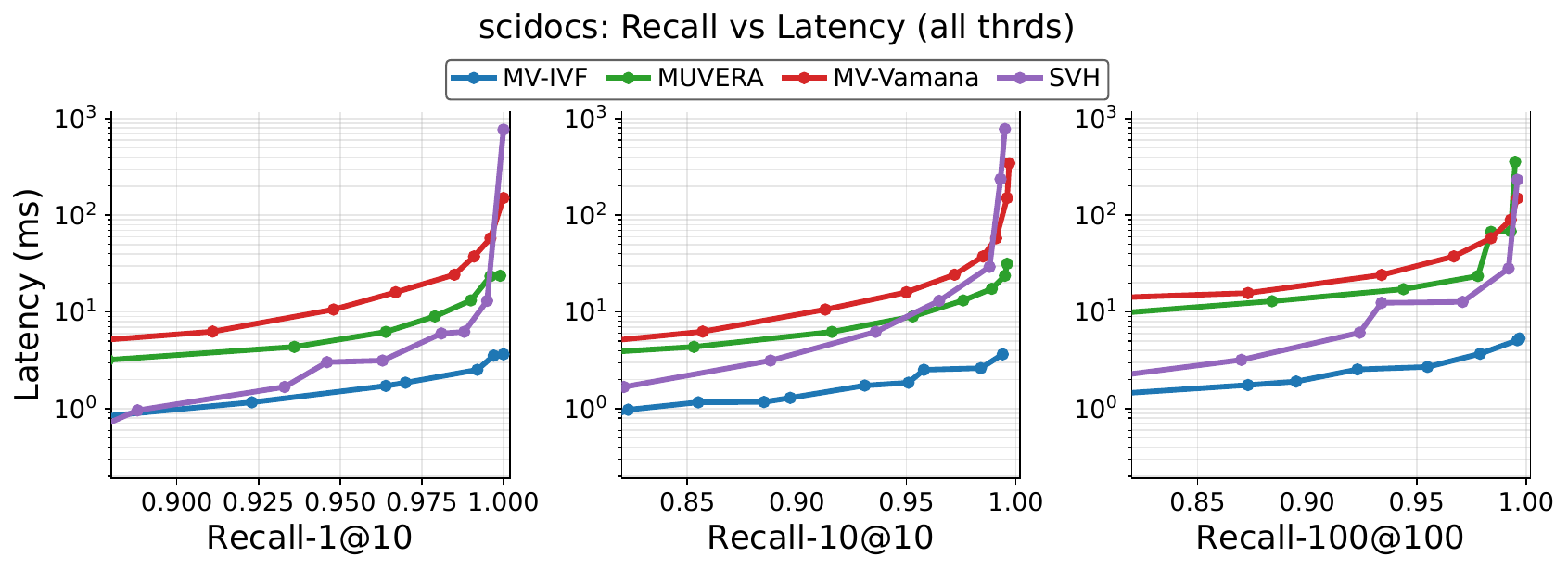}{Multi-thread latency}{scidocs}{fig:app-multi-latency-scidocs}
\appendixevaluationplot{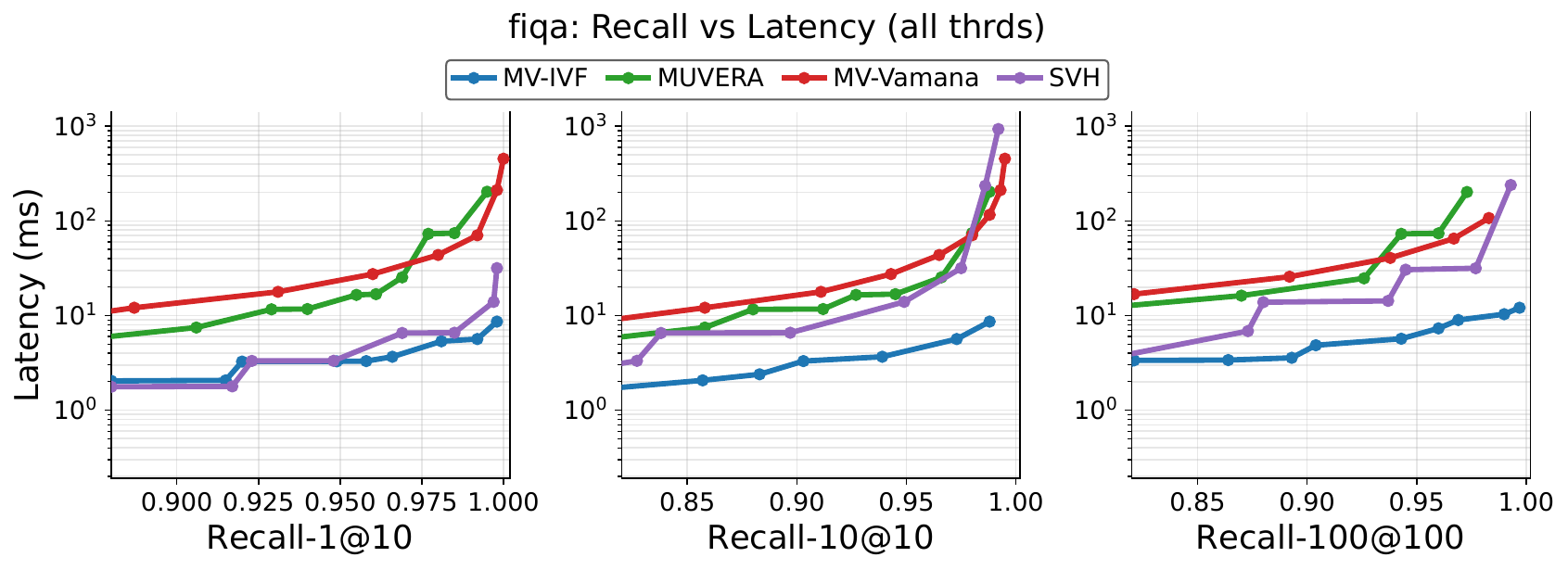}{Multi-thread latency}{fiqa}{fig:app-multi-latency-fiqa}
\appendixevaluationplot{multi_latency_quora.pdf}{Multi-thread latency}{quora}{fig:app-multi-latency-quora}

\end{document}